\documentclass[twocolumn,showpacs,aps,amsmath,amsfonts]{revtex4}

\usepackage{graphicx}
\usepackage{epic,eepic}
\usepackage{amsthm}
\usepackage{hyperref}

\newtheorem{theorem}{Theorem}

\newtheorem{lemma}[theorem]{Lemma}
\newtheorem{corollary}[theorem]{Corollary}
\newtheorem{definition}[theorem]{Definition}
\newtheorem*{proto1}{Protocol 1}

\newcommand{\lp}{\left}
\newcommand{\rp}{\right}
\newcommand{\be}{\begin{eqnarray}}
\newcommand{\ee}{\end{eqnarray}}
\newcommand{\beq}{\begin{equation}}
\newcommand{\eeq}{\end{equation}}
\newcommand{\ba}{\begin{array}}
\newcommand{\ea}{\end{array}}

\newcommand{\ket}[1]{{|{#1}\rangle}}
\newcommand{\bra}[1]{{\langle{#1}|}}

\newcommand{\mypmatrix}[1]{\begin{pmatrix}#1\end{pmatrix}}

\DeclareMathOperator{\Tr}{Tr}

\DeclareMathOperator{\Exp}{Exp}

\newcommand{\G}{\mathcal{G}}
\newcommand{\HH}{\mathcal{H}}
\newcommand{\A}{\mathcal{A}}
\newcommand{\B}{\mathcal{B}}
\newcommand{\PP}{\mathcal{P}}
\newcommand{\W}{\mathcal{W}}
\newcommand{\Z}{\mathcal{Z}}
\newcommand{\Hil}{H}

\begin{document}


\title{A large family of quantum weak coin-flipping protocols}

\author{Carlos Mochon}
\email{carlosm@theory.caltech.edu}
\affiliation{Institute for Quantum Information, 
California Institute of Technology,
Pasadena, CA 91125, USA}

\date{December 25, 2005}

\begin{abstract}
Each classical public-coin protocol for coin flipping is naturally
associated with a quantum protocol for weak coin flipping. The quantum
protocol is obtained by replacing classical randomness with quantum
entanglement and by adding a cheat detection test in the last round that
verifies the integrity of this entanglement. The set of such protocols
defines a family which contains the protocol with bias $0.192$ previously
found by the author, as well as protocols with bias as low as $1/6$
described herein. The family is analyzed by identifying a set of optimal
protocols for every number of messages. In the end, tight lower bounds for
the bias are obtained which prove that $1/6$ is optimal for all protocols
within the family.
\end{abstract}



\maketitle

\section{Introduction}

Quantum weak coin flipping is a two party quantum protocol for agreeing on
a random classical bit, where Alice wants outcome zero and Bob wants
outcome one. Its main constraint is that a cheating player should not be
able to bias the coin in their favor by more than some parameter
$\epsilon$.

Previous work by the same author \cite{me2004} has shown that there exists
a quantum weak coin-flipping protocol with bias $\epsilon=0.192$, that is,
such that neither player can win by cheating with a probability greater
than $0.692$. The protocol with bias $0.192$ was a generalization of the
one by Spekkens and Rudolph \cite{Spekkens2002} which achieved a bias of
$1/\sqrt{2}-1/2\simeq 0.207$. Both belong to a large family of quantum weak
coin-flipping protocols that are based on a set of classical games
involving public coins.

The purpose of this paper is to study this large family of protocols for
quantum weak coin flipping. In particular, we will prove that the optimal
protocol in this family has a bias of $1/6$, though such a bias can only be
reached in the limit of arbitrarily large messages. Because our lower bound
analysis is constructive, we shall give explicit descriptions of protocols
with biases that are arbitrarily close to $1/6$.

The protocols with bias of $1/\sqrt{2}-1/2$ was originally described in
Ref.~\cite{Spekkens2002} as part of a different family of protocols for
quantum weak coin flipping, all of which involved three messages. Lower
bounds for this family were obtained by Ambainis \cite{Ambainis2002bis}, 
which proved that the $\epsilon=1/\sqrt{2}-1/2$ protocol was optimal within
the family. Though our family does not contain every protocol in the
Spekkens and Rudolph family, it does contain its optimal protocol.

The best lower bound currently known that applies to all weak coin-flipping
protocols is by Ambainis \cite{Ambainis2002} and states that the number of
messages must grow at least as $\Omega(\log \log \frac{1}{\epsilon})$.
Ambainis' result rules out attaining an arbitrarily small bias with a fixed
number of messages, thus the importance of looking at protocols with
arbitrarily large number of messages. We believe that our result is the
first of its kind in lower bounding the bias of a large family of protocols
that includes instances with every number of messages.

Other important work related to quantum weak coin flipping includes
Refs.~\cite{Lo:1998pn, Goldenberg:1998bx, Spekkens2001, me2003-2,
Spekkens2003, ker-nayak} among others. Also related are the results on
quantum strong coin flipping (a variant where ideally neither player should
be allowed to bias the coin in either direction). The best known protocol
for strong coin flipping has a bias of $1/4$
\cite{Ambainis2002,Spekkens2001} whereas Kitaev \cite{Kitaev} has proven a
lower bound of $1/\sqrt{2}-1/2$ for the optimal bias.

Before proceeding we shall give a working definition of quantum weak coin
flipping as a quantum communication protocol where two parties (Alice and
Bob) start off unentangled and then exchange a series of sequential quantum
messages after which they must each output a single classical bit. Their
outputs are required to satisfy the following constraints
\begin{itemize}
\item If Alice and Bob both follow the protocol their outputs must always
agree. Furthermore, the probability that Alice wins (i.e., both parties
output zero) is given by $P_A$ whereas the probability that Bob wins (i.e.,
both parties output one) is given by $P_B=1-P_A$.
\item If Alice is honest (i.e., follows the protocol), then independent of 
Bob's actions, Alice will not output one with a probability greater than
$P_B^*$.
\item Similarly, if Bob is honest and Alice is dishonest, Bob will not
output zero with a probability greater than $P_A^*$.
\end{itemize}

\noindent
The only security assumption for the above protocol is that a cheating
player cannot directly affect the qubits in their opponent's laboratory;
that is, we desire protocols with information-theoretic security.

The parameters $P_A$, $P_A^*$ and $P_B^*$ will be used to describe a
coin-flipping protocol. Obviously, we'd like to make $P_A^*$ and $P_B^*$ as
small as possible. For simplicity, the merit of a coin-flipping protocol is
often quoted by specifying the bias $\epsilon=\max(P_A^*,P_B^*)-1/2$.

Note that whereas the usual definition of coin flipping requires
$P_A=P_B=1/2$, we will allow in this paper any value of $P_A\in[0,1]$.
This will allow us to derive a set of tradeoff curves for $P_A^*$ versus
$P_B^*$. 

The rest of the paper is organized as follows: Sec.~\ref{sec:not} describes
some of our notation concerning tree variables, and will introduce the
theorem relating classical coin games to quantum protocols for weak coin
flipping. The theorem, which is a generalization of the work in
Ref.~\cite{me2004}, is proven in Appendix~\ref{sec:theorem}.  Though the
full description of the quantum protocol is only given in the appendix, a
brief description is presented at the end of Sec.~\ref{sec:not}. 

The main new results of the paper are presented in the two subsequent
sections: the proof of lower bounds for the bias in Sec.~\ref{sec:lower}
and the description of matching protocols in Sec.~\ref{sec:upper}.

We also include in Appendix~\ref{sec:192} an analytic derivation of the
bias $\epsilon=0.192$ of Ref.~\cite{me2004} which was originally found
using numerical techniques. Though the result itself has been superseded by
the protocols with bias $1/6$, we include the derivation because it uses a
fairly different set of techniques that could potentially be useful
elsewhere.

\section{\label{sec:not}Notation}

Throughout this paper we shall make ample use of binary trees. All trees
henceforth will be composed exclusively of binary nodes and leaves, and the
leaves will all be located at the same depth.

The nodes of a tree will be labeled by binary strings so that the leftmost
node at depth $k$ gets labeled by $k$ zeroes, and the rest will equal one
plus the binary value of the node to their left (keeping the number of
digits constant). The root node will be denoted by the letter $r$, which
will behave as the empty string so that $x=r$ implies $x0=0$ and
$x1=1$. With these conventions the left descendant of node $x$ is $x0$ and
the right descendant is $x1$. We define $|x|$ as the length of the binary
string $x$, which also corresponds to the depth of node $x$.

\begin{figure}[tb]
\begin{center}
\setlength{\unitlength}{0.00075in}
\begin{picture}(4292,2400)(0,-10)
\path(600,525)(150,525)(150,825)
	(600,825)(600,525)
\path(225,525)(75,225)
\path(525,525)(675,225)
\path(1800,525)(1350,525)(1350,825)
	(1800,825)(1800,525)
\path(1425,525)(1275,225)
\path(1725,525)(1875,225)
\path(3000,525)(2550,525)(2550,825)
	(3000,825)(3000,525)
\path(2625,525)(2475,225)
\path(2925,525)(3075,225)
\path(4200,525)(3750,525)(3750,825)
	(4200,825)(4200,525)
\path(3825,525)(3675,225)
\path(4125,525)(4275,225)
\path(1200,1125)(750,1125)(750,1425)
	(1200,1425)(1200,1125)
\path(3600,1125)(3150,1125)(3150,1425)
	(3600,1425)(3600,1125)
\path(375,825)(825,1125)
\path(1125,1125)(1575,825)
\path(2775,825)(3225,1125)
\path(3525,1125)(3975,825)
\path(2400,1725)(1950,1725)(1950,2025)
	(2400,2025)(2400,1725)
\path(975,1425)(2025,1725)
\path(2325,1725)(3375,1425)
\put(2175,1875){\makebox(0,0)[c]{$\G_{r}$}}
\put(975,1270){\makebox(0,0)[c]{$\G_0$}}
\put(3375,1270){\makebox(0,0)[c]{$\G_1$}}
\put(375,675){\makebox(0,0)[c]{$\G_{00}$}}
\put(1575,675){\makebox(0,0)[c]{$\G_{01}$}}
\put(2775,675){\makebox(0,0)[c]{$\G_{10}$}}
\put(3975,675){\makebox(0,0)[c]{$\G_{11}$}}
\put(75,100){\makebox(0,0)[c]{$\G_{000}$}}
\put(675,100){\makebox(0,0)[c]{$\G_{001}$}}
\put(1275,100){\makebox(0,0)[c]{$\G_{010}$}}
\put(1875,100){\makebox(0,0)[c]{$\G_{011}$}}
\put(2475,100){\makebox(0,0)[c]{$\G_{100}$}}
\put(3075,100){\makebox(0,0)[c]{$\G_{101}$}}
\put(3675,100){\makebox(0,0)[c]{$\G_{110}$}}
\put(4275,100){\makebox(0,0)[c]{$\G_{111}$}}
\end{picture}
\end{center}
\caption{A depth 3 binary tree.}
\label{fig:not}
\end{figure}
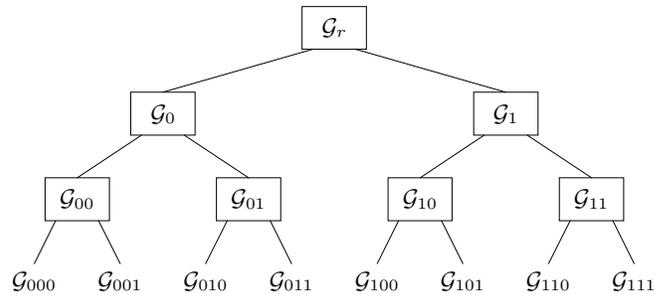

In this paper we shall use calligraphic fonts, such as $\G$, to denote an
assignment of a number or expression to each node of a binary tree. Given
an assignment $\G$, the value of node $x$ will be $\G_x$. Most of our
notation is summarized by Fig.~\ref{fig:not}. Note that, though we shall
always be working with trees of fixed finite depth, we shall usually leave
the depth implicit.

We define an $n$-Coin-Game as an assignment $\G$ to a depth $n$ binary tree
such that $\G_x\in[0,1]$ for all $x$ and $\G_x\in\{0,1\}$ for all leaves
(i.e., for all $x$ such that $|x|=n$). To each $n$-Coin-Game, $\G$, we can
associate a classical $n$-message public-coin coin-flipping protocol as
follows: The state of the protocol at each step will be described by a node
in the tree, and this information will be kept by both Alice and Bob. The
game begins at the root node and proceeds downward until reaching a leaf
node. If the current node $x$ is a binary node of even depth, then Alice
chooses which path to follow and announces the choice to Bob. This is done
probabilistically, by announcing the outcome of a public coin with
bias $\G_x$, so that Alice chooses the left path with probability $\G_x$
and the right path with probability $1-\G_x$. The same mechanism occurs at
odd binary nodes, except that Bob is responsible for choosing the direction
and announcing it to Alice. The game ends when arriving at a leaf node $x$,
in which case Alice wins if $\G_x=0$ and Bob wins if $\G_x=1$.

Note that we do not require that the coin-flip be fair when both Alice and
Bob are honest. Given an $n$-Coin-Game $\G$, we can define $\HH$ on a tree
of the same depth by the equations:
\be
\HH_x = 
\begin{cases} 
\G_x & \text{if $|x|=n$,}\\ 
\G_x \HH_{x0} + (1-\G_x) \HH_{x1} & \text{if $|x|<n$.} 
\end{cases}
\label{eq:h}
\ee

\noindent
The value of $\HH_x$ indicates the conditional probability that Bob would
win given that the game arrived at node $x$, assuming both players
play honestly. The value of $\HH_{r}$ is Bob's probability of winning for
an honest game, which is clearly bounded between 0 and 1.

For each $n$-Coin-Game $\G$, we also define $\A$ and $\B$ on a tree of the
same depth by the equations:
\be
\A_x = 
\begin{cases} 
1-\G_x & \text{for $|x|=n$,}\\ 
\G_x \A_{x0}^2 + (1-\G_x) \A_{x1}^2 & \text{$|x|$ even, $|x|<n$,} \\
\G_x \sqrt{\A_{x0}} + (1-\G_x) \sqrt{\A_{x1}} & \text{$|x|$ odd, $|x|<n$,} \\
\end{cases}\nonumber\\\nonumber
\B_x = 
\begin{cases} 
\G_x & \text{for $|x|=n$,}\\ 
\G_x \sqrt{\B_{x0}} + (1-\G_x) \sqrt{\B_{x1}} & \text{$|x|$ even, $|x|<n$,} \\
\G_x \B_{x0}^2 + (1-\G_x) \B_{x1}^2 & \text{$|x|$ odd, $|x|<n$.} \\
\end{cases}
\\*\label{eq:ab}
\ee

The importance of these quantities is given by the following theorem:

\begin{theorem}
For each $n$-Coin-Game, $\G$, there exists an $(n+1)$-message quantum weak
coin-flipping protocol such that
\be
P_A P_A^* &=& \A_{r},\\
P_B P_B^* &=& \B_{r}^2,
\ee
\noindent
and the honest probabilities of winning are
\be
P_A = (1-P_B) = (1-\HH_{r}),
\ee
where $\A$, $\B$ and $\HH$ are defined in terms of $\G$ by
Eqs.~(\ref{eq:h},\ref{eq:ab}).
\label{theorem:proto}
\end{theorem}

\subsection*{The quantum protocol}

In this section we shall give a brief approximate description of the
quantum protocol, which should provide the needed intuition. The full
description of the protocol is contained in Appendix~\ref{sec:theorem}
along with the proof of the above theorem. A simpler version of the
protocol also appears in Ref.~\cite{me2004}.

The basic idea is to take the classical public-coin protocol associated with 
an $n$-Coin-Game, $\G$, replace the classical randomness with quantum
entanglement, and then add a cheat detection step.

Classical shared randomness can be replaced by quantum entanglement using
states of the form
\be
\sqrt{a}\ket{0}\otimes\ket{0}+\sqrt{1-a}\ket{1}\otimes\ket{1},
\ee
where one qubit belongs to Alice and one to Bob. The randomness can be
extracted at any time by measuring both qubits in the computational
basis.

In the classical protocol associated with $\G$ described above, Alice and
Bob slowly built up a shared random string.  After the first $k$ messages
they shared a random $k$-bit string, where string $x$ has probability
$\PP_x$ (the formal definition of $\PP$ is given in Eq.~(\ref{eq:p})). The
quantum protocol is constructed so that, after $k$ messages, Alice and Bob
share the state
\be
\ket{\psi_k}=\sum_{\substack{x\cr |x|=k}} \sqrt{\PP_x} \ket{x}\otimes\ket{x}.
\ee
In the classical protocol, the sender of the message (Alice for odd
messages and Bob for even messages) has control over its content and hence
the ability to cheat at that step, whereas the other player has no control
over the given step. In the quantum protocol the same structure is
maintained. The basic step to go from a $k$-bit string to a $k+1$-bit
string is for the message sender to append two qubits in the zero state,
then apply a controlled unitary on the two qubits with the other $k$ bits
as control, and finally to send one of the qubits to the other player:
\be
\ket{\psi_k}&\longrightarrow&
\ket{\psi_k}\otimes \ket{00}
\\\nonumber
&\longrightarrow&
\sum_{\substack{x\cr|x|=k}}
\sqrt{\PP_x} \ket{x}_A\otimes\ket{x}_B
\\\nonumber
&&\qquad \quad
\otimes \lp(\sqrt{\G_{x}}\ket{00}+\sqrt{1-\G_{x}}\ket{11}\rp)
\\\nonumber
&\longrightarrow&
\sum_{\substack{x\cr|x|=k}}\sum_{i\in\{0,1\}}
\sqrt{\PP_{xi}} \ket{xi}_A\otimes\ket{xi}_B
= \ket{\psi_{k+1}}.
\ee

After $n$ messages, at the end of the classical protocol, Alice and Bob
share an $n$-bit string, which determines the coin outcome based on the
value of the corresponding leaf of $\G$. In the quantum protocol, they do
the equivalent measurement, but using a two outcome POVM so that most of the
entanglement is preserved after the measurement. This allows a cheat
detection step to be appended to the end of the protocol as follows: the
winner of the coin-flip based on the POVM must send over all of their
qubits to the other player for inspection. The other player will end up
with a pure state and can do a projection onto the final state and its
complement. If the latter result is obtained, then cheating is detected and
the losing player can declare victory, otherwise that player acknowledges
defeat. In either case, the first player always declares victory.

Note that, though it is possible for both players to declare victory at the
same time, this can only occur if one of them was cheating, and in such
cases we always expect the cheating player to declare victory anyway.

The rest of this paper contains the analysis of the family of protocols,
which will identify the protocol with bias of $1/6$ and prove that it is
optimal within the family.

\section{\label{sec:lower}Lower bounds on the bias}

In this section we shall derive lower bounds for the set of $P_A^*$ and
$P_B^*$ that can be achieved with quantum protocols based on $n$-Coin-Games
as defined in Theorem~\ref{theorem:proto}.

\begin{definition}
For $n\in\mathbb{Z}^+$, define the set $\Lambda_n\subset\mathbb{R}^2$ 
so that $(A,B)\in\Lambda_n$ if and only if there exists an $n$-Coin-Game,
$\G$, with $A=\A_r$ and $B=\B_r$ and $\A$ and $\B$ defined in
terms of $\G$ by Eq.~(\ref{eq:ab}).
\end{definition}

For each $(A,B)\in\Lambda_n$ there exists an $(n+1)$-message quantum
coin-flipping protocol such that $P_A P_A^*=A$ and $P_B P_B^*=B^2$.
Furthermore, if $(P_A P_A^*, \sqrt{P_B P_B^*})\notin\Lambda_n$ then there
is no protocol built out of a $n$-Coin-Game that achieves $P_A$, $P_A^*$
and $P_B^*$. However, it is not true that $(P_A P_A^*, \sqrt{P_B
P_B^*})\in\Lambda_n$ implies the existence of a protocol with those
parameters. For example, $(0.3531,\sqrt{0.3531})\in\Lambda_2$ because there
exists a $3$-message protocol with $P_A\simeq 0.515$, $P_A^*\simeq 0.686$,
$P_B^*\simeq 0.728$, however there are no $3$-message protocols with
$P_A=P_B=1/2$ and $P_A^*=P_B^*\simeq 2*0.353=0.706$. The optimal symmetric
3-message protocol is the one by Spekkens and Rudolph \cite{Spekkens2002}
with $P_A^*=P_B^*=1/\sqrt{2}=0.707$. Though it would be preferable to
study the set of achievable triplets $(\A_r,\B_r,\HH_r)$, the
sets $\Lambda_n$ are easier to analyze and in the limit
$n\rightarrow\infty$ will provide us with interesting bounds.

We begin the study of the sets $\Lambda_n$ by showing that they can be
obtained inductively:

\begin{lemma}
The set $\Lambda_n$ is the convex combination of pairs of points from the set 
$\{(B^2,\sqrt{A})\,|\,(A,B)\in\Lambda_{n-1}\}$.
\label{lemma:lambda}
\end{lemma}

\begin{proof}
Given an $n$-Coin-Game, $\G$, define the variable
$\gamma\equiv\G_r\in[0,1]$ and the two $(n-1)$-Coin-Games $\G^{(0)}$ and
$\G^{(1)}$ by
\be
\G^{(i)}_x=
\begin{cases} 
1-\G_{ix} & \ \ \text{for $\ |x|=n-1$,}\\
\G_{ix} & \ \ \text{for $\ |x|<n-1$,}
\end{cases}
\ee
\noindent
for $i=0,1$. There is a natural isomorphism between $\G$ and the triplet
$\gamma,\G^{(0)},\G^{(1)}$.

Furthermore define $\A^{(i)}$ and $\B^{(i)}$ in terms of $\G^{(i)}$ in the
usual way. Note that $\A^{(i)}$ and $\B^{(i)}$ are not the left and right
branches of $\A$ and $\B$ defined from $\G$ but rather $\A^{(i)}_x=\B_{ix}$
and $\B^{(i)}_x=\A_{ix}$. Therefore
\be
\A_r &=& \gamma \lp(\B_r^{(0)}\rp)^2 + (1-\gamma) \lp(\B_r^{(1)}\rp)^2,\\
\B_r &=& \gamma \sqrt{\A_r^{(0)}} + (1-\gamma) \sqrt{\A_r^{(1)}}.
\ee
\end{proof}

The set $\Lambda_1$ is fairly simple and corresponds to the convex
combinations of the two points $(1,0)$ and $(0,1)$, which could be thought
of as comprising $\Lambda_0$. Using $\Lambda_1$ and the above lemma we can
prove two simple properties of the sets $\Lambda_n$:

\begin{enumerate}
\item $(0,1)\in\Lambda_n$ and $(1,0)\in\Lambda_n$ for all $n$.
\item $\Lambda_n\subset [0,1]\times[0,1]$ for all $n$.
\end{enumerate}

\noindent
Both properties are clearly true for $\Lambda_1$. By induction $(0,1)\in
\Lambda_{n-1}$ and $(1,0)\in\Lambda_{n-1}$ implies that  
$(1^2,\sqrt 0)$ and $(0^2,\sqrt 1)$ are in $\Lambda_n$. Similarly, if
$(A,B)\in\Lambda_{n-1}$ implies $A\in[0,1]$ and $B\in[0,1]$, then
$(B^2,\sqrt{A})\in[0,1]\times[0,1]$ and so are convex combinations of such
points.

\begin{figure}[tb]
\includegraphics[scale=.8]{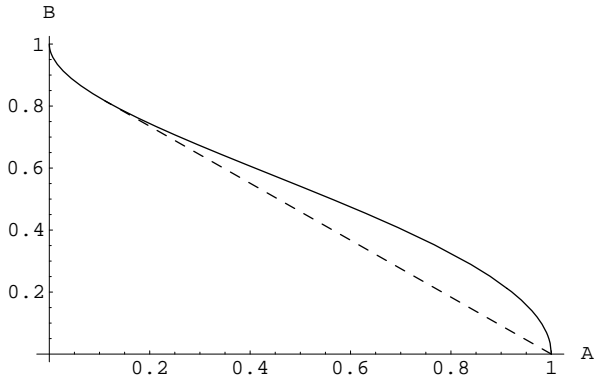}
\caption{The curve $(t^2,\sqrt{1-t})$ for $t\in[0,1]$. The convex hull
of the curve is the region $\Lambda_2$, with the dashed line serving as
lower boundary.}
\label{fig:n2plot}
\end{figure}

The first non-trivial set is $\Lambda_2$ which is the convex combination
of the points on the curve $(t^2,\sqrt{1-t})$ for $t\in[0,1]$. The curve is
plotted in Fig.~\ref{fig:n2plot}. The dotted line marks the lower boundary
of its convex hull which can be achieved using convex combinations of
two points (the rest of the lower boundary of the convex hull is simply
the curve itself).

Rather than keeping track of the sets $\Lambda_n$, it will be simpler to
study exclusively their lower boundary, which will be curves connecting the
points $(1,0)$ and $(0,1)$. All the optimal protocols will live on these
curves, and all points below the curves will be unattainable.
To formalize the notion of lower boundary we associate to every function
$f(z):[0,1]\rightarrow[0,1]$ the following sets:
\be
f^+ &=& \{(z,w)\,|\,z\in[0,1],f(z)<w\leq 1\},\\
f^= &=& \{(z,w)\,|\,z\in[0,1],f(z)=w\},\\
f^- &=& \{(z,w)\,|\,z\in[0,1],f(z)>w\geq 0\}.
\ee

Returning to the case of $\Lambda_2$ and
Fig.~\ref{fig:n2plot}, we see that the lower boundary follows
the original curve $\sqrt{1-\sqrt{z}}$ between $(1,0)$ and some point which
we shall call $(\alpha_2,\beta_2)$. It then turns into a straight line
connecting the point $(\alpha_2,\beta_2)$ to the point $(0,1)$. The point 
$(\alpha_2,\beta_2)$ can be found by calculating the slope of the line
connecting each point to $(0,1)$ and choosing the point that achieves the
maximum. 

In fact, all of the lower boundaries will have this form. Define for $n >
1$
\be
f_n(z)=\begin{cases} 
\sqrt{1-\lp(\frac{1-\beta_n^2}{\sqrt{\alpha_n}}\rp)\sqrt{z}}
& \text{for $z\in[0,\alpha_n]$,}\\ 
\frac{\beta_n}{1-\alpha_n}(1-z)
& \text{for $z\in[\alpha_n,1]$,} 
\end{cases}
\label{eq:f}
\ee

\noindent
where
\be
\alpha_n = \frac{n-1}{3(n+1)},\quad\quad\beta_n=\sqrt{\frac{n+2}{3n}}.
\ee

\noindent
For the case $n=1$ we define $f_1(z)=1-z$, which is the limit of $f_n$ as
$n\rightarrow 1$. Because $\alpha_n\in(0,1)$ and $\beta_n\in(0,1)$ for all
$n>1$, the functions satisfy $f_n(z)\in[0,1]$ for all $z\in[0,1]$. These
functions are also the lower boundaries of convex regions:

\begin{lemma}
For all $n\geq 1$, the function $f_n$ is strictly decreasing, and the
region $f_n^=\cup f_n^+$ is convex.
\label{lemma:f}
\end{lemma}

\begin{proof}
The case of $n=1$ is trivial. For $n>1$ we have
\be
f_n'(z)=\begin{cases} 
-\frac{1-\beta_n^2}{4\sqrt{\alpha_n}\sqrt{z}f_n(z)}
& \text{for $z\in[0,\alpha_n]$,}\\ 
-\frac{\beta_n}{1-\alpha_n}
& \text{for $z\in[\alpha_n,1]$,} 
\end{cases}
\ee
\noindent
which is well defined and negative on $(0,1]$. For $z$ near zero,
$f(z)\simeq 1-(1-\beta_n^2)/(2\sqrt{\alpha_n})\sqrt{z}$, therefore $f(z)$
is also strictly decreasing at $z=0$.

The derivative is also continuous on $(0,1]$ because at $z=\alpha_n$ we
have
\be
\frac{\beta_n}{1-\alpha_n} = \frac{\sqrt{3}(n+1)}{2\sqrt{n(n+2)}}
= \frac{1-\beta_n^2}{4\alpha_n\beta_n}.
\ee
Furthermore, in the region $(0,\alpha_n)$, the second derivative is
\be
f_n''(z) &=& f'_n(z)\lp[ -\frac{1}{2z}-\frac{f_n'(z)}{f_n(z)}\rp]
\label{eq:fpp}
\\\nonumber
&=& \frac{-f_n'(z)}{4z\sqrt{\alpha_n}f_n^2(z)}
\lp[2\sqrt{\alpha_n}-3(1-\beta_n^2)\sqrt{z}\rp]>0 
\ee
\noindent
where the inequality holds because $3(1-\beta_n^2)<2$. Therefore $f'_n(z)$
is monotonically increasing on $(0,1]$, and the region above $f_n(z)$ in
this interval is convex. The point $(0,1)$ can be included because the
closure of a convex set is convex.
\end{proof}

We are now ready to prove the main lemma of this section.

\begin{lemma}
For all $n\in\mathbb{Z}^+$, $\Lambda_n\subset f_n^=\cup f_n^+$ and 
$f_n^=\subset\Lambda_n$.
\end{lemma}

\begin{proof}
The statement is clearly true for $n=1$ since $\Lambda_1=f_1^=$. We will
prove the rest of the cases inductively. Assume the theorem holds for
$\Lambda_n$, which implies that $(z,f_n(z))\in\Lambda_n$ for all $z$.  By
Lemma~\ref{lemma:lambda} we have that $(f^2_n(z),\sqrt{z})\in\Lambda_{n+1}$
for all $z\in[0,1]$ and so are convex combinations of pairs of such
points. The curve parametrized by $(f^2_n(z),\sqrt{z})$ can also be
described by the points $(w,g_n(w))$ for
\be
g_{n}(w)=\begin{cases} 
\sqrt{1-\lp(\frac{1-\alpha_n}{\beta_n}\rp)\sqrt{w}}
& \text{for $w\in[0,\beta_n^2]$,}\\ 
\frac{\sqrt{\alpha_n}}{1-\beta_n^2}(1-w)
& \text{for $w\in[\beta_n^2,1]$.} 
\end{cases}
\label{eq:g}
\ee
Note how under the map $(x,y)\rightarrow(y^2,\sqrt{x})$ the straight line
turns into a curve, and the curve turns into a straight
line. Furthermore, because of the exchange of $x$ and $y$, the straight line
ends up on the right-hand side.

The pattern of points $\alpha_n$ and $\beta_n$, in addition to guaranteeing
that the region above $f_n(z)$ is convex, also satisfies the recursion
relation
\be
\frac{1-\alpha_n}{\beta_n} = \frac{2\sqrt{n(n+2)}}{\sqrt{3}(n+1)} =
\frac{1-\beta_{n+1}^2}{\sqrt{\alpha_{n+1}}}
\ee
and therefore $g_n(z)=f_{n+1}(z)$ in the region $[0,\alpha_{n+1}]$ (since
$\alpha_{n+1}\leq1/3\leq\beta_n^2$). Pictorially, the curve $g_n^=$ is like
the curve $f_{n+1}^=$, except that the straight line intersects the curve
somewhat to the right, and hence the region above $g_n^=$ is not
convex. Its convex hull will give us the region above the curve
$f_{n+1}^=$.

Thus far we have shown $g_n^=\subset\Lambda_{n+1}$, as are convex
combinations of pairs of points on the curve $g_n^=$. Because
$g_n^==f_{n+1}^=$ in the region $[0,\alpha_{n+1}]$ we know that this
segment of the curve is in $\Lambda_{n+1}$. The rest of the curve
$f_{n+1}^=$ is simply the convex combination of the points
$(\alpha_{n+1},\beta_{n+1})$ and $(1,0)$ both of which are in $g_n^=$. We
have therefore proven the second part of the lemma:
$f_{n+1}^=\subset\Lambda_{n+1}$.

We now intend to prove that $g_n(z)\geq f_{n+1}(z)$ for all
$z\in[0,1]$. The statement is clearly true in the region $[0,\alpha_{n+1}]$
where both are equal. In the region $[\beta_n^2,1]$ it is also true because
both functions are straight lines ending in $(1,0)$, and the starting point
of the lines are $g_n(\beta_n^2)=\sqrt{\alpha_n}$ and $f_{n+1}(\beta_n^2) =
\beta_{n+1}(1-\beta_n^2)/(1-\alpha_{n+1})$. The inequality 
$f_{n+1}(\beta_n^2)\geq g_n(\beta_n^2)$ can be proven by checking that
$[f_{n+1}(\beta_n^2)/g_n(\beta_n^2)]^2-1= -4/[n^2(n+3)]\leq 0$ for
$n\geq1$. Finally, in the region $[\alpha_{n+1},\beta_n^2]$ the functions
$f_{n+1}(z)$ and $g_{n}(z)$ start off at the same point, with the same
derivative, but $f_{n+1}''(z)=0$ in this region whereas $g_{n}''(z)$
initially is positive, and has only one zero in the region, which can be
checked as in Eq.~(\ref{eq:fpp}). If the curve $g_{n}$ were to cross the
curve $f_{n+1}$ at any point in this region, then it would have to end
below it. However, we already argued that $g_n(\beta_n^2)\geq
f_{n+1}(\beta_n^2)$ and therefore the curve $g_n^=$ must lie above the
curve $f_{n+1}^=$ in the middle region as well.

So far we have shown that $(g_{n}^=\cup g_{n}^+) \subset (f_{n+1}^=\cup
f_{n+1}^+)$. By the induction assumption, $\Lambda_n\subset f_{n}^=\cup
f_{n}^+$. Under the map $(x,y) \rightarrow (y^2,\sqrt{x})$, the region
$f_{n}^=\cup f_{n}^+$ maps into the region to the right of the curve
$g_{n}^=$, which also equals the region $g_{n}^=\cup g_{n}^+$ because
$g_n(z)$ is strictly decreasing, $g_n(1)=0$, and $g_n(0)=1$.  Finally,
using Lemma~\ref{lemma:lambda} we know that $\Lambda_{n+1}$ is contained in
the convex combination of points in $g_{n}^=\cup g_{n}^+$.  Because
$(g_{n}^= \cup g_{n}^+) \subset (f_{n+1}^= \cup f_{n+1}^+)$, and $f_{n+1}^=
\cup f_{n+1}^+$ is convex, we have $\Lambda_{n+1} \subset f_{n+1}^=\cup
f_{n+1}^+$.
\end{proof}

Combining the previous lemma with the definition of the sets
$\Lambda_n$, we have proven the following theorem:

\begin{theorem}
Every $(n+1)$-message quantum weak coin-flipping protocol based on an
$n$-Coin-Game satisfies 
\be
P_B P_B^* \geq f_n^2(P_A P_A^*).
\ee
\end{theorem}

\noindent
Additionally, we have the following corollary for the limit of
$n\rightarrow\infty$:

\begin{corollary}
All quantum weak coin-flipping protocols based on an $n$-Coin-Game (for
any $n\in\mathbb{Z}^+$) satisfy
\be
P_A P_A^*\leq\frac{1}{3}\quad\Longrightarrow\quad
P_B P_B^* \geq 1-2\sqrt{\frac{P_A P_A^*}{3}}\geq\frac{1}{3}\\
P_B P_B^*\leq\frac{1}{3}\quad\Longrightarrow\quad
P_A P_A^* \geq 1-2\sqrt{\frac{P_B P_B^*}{3}}\geq\frac{1}{3} 
\ee
\noindent
In particular,
\be
\max\lp(P_A P_A^*,P_B P_B^*\rp)\geq\frac{1}{3},
\ee
\noindent
and
\be
\max(P_A^*,P_B^*)\geq \frac{2}{3}\ \ \ \  \text{for $\ P_A=P_B=\frac{1}{2}$}.
\ee
\end{corollary}

\begin{proof}
The above results use the limit:
\be
f_\infty(z)=\begin{cases} 
\sqrt{1-\frac{2}{\sqrt{3}}\sqrt{z}}
& \text{for $z\in[0,\frac{1}{3}]$,}\\ 
\frac{\sqrt{3}}{2}(1-z)
& \text{for $z\in[\frac{1}{3},1]$,} 
\end{cases}
\ee
\noindent
which has the symmetry $b=f_\infty^2(a)\Rightarrow a=f_\infty^2(b)$.
\end{proof}

\section{\label{sec:upper}Optimal Protocols}

In this section we will describe protocols that match the lower bounds
derived in the previous section. In a sense, most of the work has already
been done since the proof of the previous section was constructive.
What remains undone is to explicitly construct the $n$-Coin-Games and to
calculate from them $P_A$, $P_A^*$ and $P_B^*$ (rather than only their
products).

From the discussion of the previous section we can see that the interesting
$(n+1)$-Coin-Games live on the curve $f_{n+1}^=$. The points on the
rounded part of the curve (the left segment) involve no convex combinations
of points from $n$-Coin-Games and therefore are not new (i.e., they are
protocols that can be described by a single $n$-Coin-Game with Alice's and
Bob's role reversed). The interesting points at level $n+1$ lie on the
straight segment and are the combination of the points
$(\alpha_{n+1},\beta_{n+1})$ and $(1,0)$. To understand this segment we
need to describe the $n$-Coin-Games that produce points
$(\beta_{n+1}^2,\sqrt{\alpha_{n+1}})$ and $(0,1)$. The second point
corresponds to a tree that is fairly simple: it has the value $1$ at every
leaf and the rest of the nodes are irrelevant. The $n$-Coin-Games for
$(\beta_{n+1}^2,\sqrt{\alpha_{n+1}})$ is what we shall describe next.

\begin{lemma}
For each $n\in\mathbb{Z}^+$ there is an $n$-Coin-Game, $\G^{(n)}$, such that 
\be
\label{eq:consa}
\A^{(n)}_r&=&\beta_{n+1}^2=\frac{n+3}{3(n+1)},\\
\label{eq:consb}
\B^{(n)}_r&=&\sqrt{\alpha_{n+1}}=\sqrt{\frac{n}{3(n+2)}},\\
\label{eq:consh}
\HH^{(n)}_r&=&\begin{cases} 
\frac{n}{2(n+1)}
& \text{$n$ even},\\
\frac{n+1}{2(n+2)}
& \text{$n$ odd},
\end{cases}
\ee
\noindent
with $\A^{(n)}$, $\B^{(n)}$ and $\HH^{(n)}$ defined in terms of $\G^{(n)}$
by Eqs.~(\ref{eq:h},\ref{eq:ab}). In particular, the associated quantum
weak coin-flipping protocols have:
\be
P_A(n)=&1-\HH^{(n)}_r &= \begin{cases} 
\frac{n+2}{2(n+1)}
& \text{$n$ even},\\
\frac{n+3}{2(n+2)}
& \text{$n$ odd},
\end{cases}\\
P_A^*(n)=&\frac{\A^{(n)}_r}{1-\HH^{(n)}_r} &= \begin{cases} 
\frac{2(n+3)}{3(n+2)}
& \text{$n$ even},\\
\frac{2(n+2)}{3(n+1)}
& \text{$n$ odd},
\end{cases}\\
P_B^*(n)=&\frac{\lp(\B^{(n)}_r\rp)^2}{\HH^{(n)}_r} &= \begin{cases} 
\frac{2(n+1)}{3(n+2)}
& \text{$n$ even},\\
\frac{2n}{3(n+1)}
& \text{$n$ odd}.
\end{cases}
\ee
\end{lemma}

\begin{proof}
Define the parameters
\be
\gamma_n = \frac{n}{n+2},
\ee
\noindent
which are the weights needed for the convex combinations. And let
\be
\G^{(1)}_r=\gamma_1,\ \ \ \ \G^{(1)}_0=1,\ \ \ \ \G^{(1)}_1=0,
\ee
\noindent
which leads to $\A_r^{(1)}=2/3$ and $\B_r^{(1)}=\HH_r^{(1)}=1/3$. The rest
of the Coin-Games are defined inductively:
\be
\G^{(n)}_r&=&\gamma_n,\\
\G^{(n)}_{0x}&=&\begin{cases}
1-\G^{(n-1)}_{x} & \ \text{for $\ |x|=n-1$},\\
\G^{(n-1)}_{x} & \ \text{for $\ |x|<n-1$},\\
\end{cases}\\
\G^{(n)}_{1x}&=&\begin{cases}
0 & \ \ \ \ \ \ \text{for $\ |x|=n-1$},\\
\G^{(n-1)}_{x} & \ \ \ \ \ \ \text{for $\ |x|<n-1$}.\\
\end{cases}
\ee

The values of $\G^{(n)}_{1x}$ for $|x|<n-1$ are actually irrelevant but
were chosen so that $\G^{(n)}_x=\gamma_{n-|x|}$ whenever $|x|<n-1$, and
therefore these protocols fit into the subfamily studied in
Ref.~\cite{me2004}.

The reason for inverting the value of the leaves relates to our insistence
that Alice always send the first message, which implies that the sender of
the last message alternates as $n$ is increased and correspondingly the
assignments of winning and losing for the coin outcome need to be flipped.

In fact, the pattern of the leaves is fairly simple. It is chosen so that it
depends on the parity of the location (from left to right) of the first 1
symbol in the string $x$. In the quantum protocol this translates into the
first sender of a 1 qubit being the winner of the coin-flip (assuming they
pass the cheat detection phase). 

In fact, the trees $\G^{(n)}$ would best be described by truncated trees of
the form of Fig.~\ref{fig:trunctree}. However, we shall continue using trees
with all leaves at the same depth in order to be consistent with the
previous section.

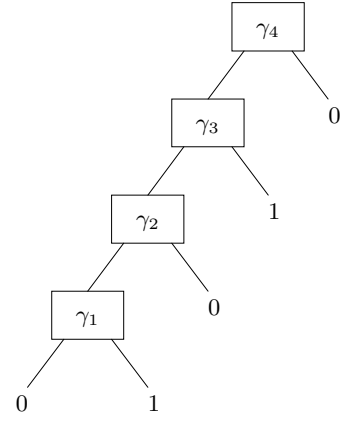
\begin{figure}[tb]
\setlength{\unitlength}{0.00083333in}
\begin{picture}(2042,2577)(0,-10)
\path(1800,2250)(1350,2250)(1350,2550)
	(1800,2550)(1800,2250)
\path(1050,1050)(600,1050)(600,1350)
	(1050,1350)(1050,1050)
\path(675,450)(225,450)(225,750)
	(675,750)(675,450)
\path(1425,1650)(975,1650)(975,1950)
	(1425,1950)(1425,1650)
\path(1425,2250)(1200,1950)
\path(1725,2250)(1950,1950)
\path(1050,1650)(825,1350)
\path(1350,1650)(1575,1350)
\path(675,1050)(450,750)
\path(975,1050)(1200,750)
\path(300,450)(75,150)
\path(600,450)(825,150)
\put(1950,1800){\makebox(0,0)[lb]{$0$}}
\put(1575,1200){\makebox(0,0)[lb]{$1$}}
\put(1200,600){\makebox(0,0)[lb]{$0$}}
\put(825,0){\makebox(0,0)[lb]{$1$}}
\put(0,0){\makebox(0,0)[lb]{$0$}}
\put(1500,2325){\makebox(0,0)[lb]{$\gamma_4$}}
\put(1125,1725){\makebox(0,0)[lb]{$\gamma_3$}}
\put(750,1125){\makebox(0,0)[lb]{$\gamma_2$}}
\put(375,525){\makebox(0,0)[lb]{$\gamma_1$}}
\end{picture}
\caption{A truncated tree equivalent to $\G^{(4)}$.}
\label{fig:trunctree}
\end{figure}

Returning to the proof of the lemma, it is easy to see that $\A_{1x}^{(n)}=1$
and $\B_{1x}^{(n)}=\HH_{1x}^{(n)}=0$ for all strings $x$. The left side of
the tree satisfies $\A_{0x}^{(n)}=\B_{x}^{(n-1)}$,
$\B_{0x}^{(n)}=\A_{x}^{(n-1)}$ and $\HH_{0x}^{(n)}=1-\HH_{x}^{(n-1)}$ for
all strings $x$. Therefore the root nodes are
\be
\A_r^{(n)} = &\gamma_n \lp(\B_r^{(n-1)}\rp)^2& + (1-\gamma_n)\ 1,\\
\B_r^{(n)} = &\gamma_n \sqrt{\A_r^{(n-1)}}& + (1-\gamma_n)\ 0,\\
\HH_r^{(n)} = &\gamma_n \lp(1-\HH_r^{(n-1)}\rp)& + (1-\gamma_n)\ 0.
\ee

\noindent
It is then straightforward to plug in the expressions as functions of $n$
for all the above parameters and check that
Eqs.~(\ref{eq:consa}--\ref{eq:consh}) are always satisfied.
\end{proof}

Interestingly, the sequence of protocols is such that $P_A$ and $P_B$ do
not change when $n$ increases from an odd integer to an even one, whereas
$P_A^*$ and $P_B^*$ do not change when $n$ increases from an even integer
to an odd one. We offer no intuition for this property. Note, however, that
for a given $n$, the associated protocol corresponds to a single point on the
surface of optimal protocols in the $3$-dimensional space of triplets
$(P_A,P_A^*,P_B^*)$ that can be achieved with $n+1$ quantum messages.

For large $n$, the sequence of protocols converges to $P_A=P_B=1/2$ and
$P_A^*=P_B^*=2/3$, yielding a protocol with bias of $1/6$. It would also be
desirable to show the existence of a sequence of protocols that converges to
the same point but such that $P_A=P_B=1/2$ for every protocol in the
sequence. This can be easily accomplished by choosing, for each $n$, the
point along the curve $f_n^=$ that has $\HH_r=1/2$. In the Coin-Game
language we need to modify the top coin $\G_r$, and we therefore introduce
a new sequence of Coin-Games ${\G'}^{(n)}$ defined as
\be
{\G'_x}^{(n)}=\begin{cases}
1/\lp(2-2\HH_r^{(n-1)}\rp)&x=r,\\
\G^{(n)}_x&\text{otherwise.}
\end{cases}
\label{eq:optg}
\ee
\noindent
For simplicity, we will concentrate on the case when $n$ is even so that:
\be
{\A'_r}^{(n)} &=& \frac{n+1}{n+2} \lp(\B_r^{(n-1)}\rp)^2 + \frac{1}{n+2},\\
{\B'_r}^{(n)} &=& \frac{n+1}{n+2} \sqrt{\A_r^{(n-1)}},\\
{\HH'_r}^{(n)} &=& \frac{n+1}{n+2} (1-\HH_r^{(n-1)})=\frac{1}{2}
\ee

\noindent
and the associated probabilities of winning by cheating are
\be
P_A^*(n)'=& 2 {\A'_r}^{(n)} &= \frac{2}{3},\\ 
\label{eq:optPB}
P_B^*(n)'=&2 \lp({\B'_r}^{(n)}\rp)^2 &= \frac{2}{3}\, \frac{(n+1)^2}{n(n+2)}.
\ee

\noindent
That is, we have identified a nice sequence of quantum protocols with $n+1$
messages (for $n$ even) where $P_A=P_B=1/2$ and $P_A^*=2/3$ are all fixed
and $P_B^*$ decreases from $3/4$ to $2/3$. Of course, the case $n=2$
belongs to the family studied by Spekkens and Rudolph \cite{Spekkens2002}
and satisfies $P_A^* P_B^* =1/2$.

As discussed in the introduction to the previous section, the above
protocols are optimal in the following sense: to decrease one of $P_A^*$ or
$P_B^*$ while keeping the number of messages fixed, we would have to
increase the other parameter. However, the protocols are not optimal in the
sense that they minimize the bias $\epsilon=\max(P_A^*,P_B^*)-1/2$ for a
fixed number of messages. Only in the limit of infinite messages is the
bias of the above protocols optimal. 

Thus far, we have identified the point $(1/3,\sqrt{1/3})\in f_\infty^=$ as
a protocol with $P_A=1/2$ and $P_A^*=P_B^*=2/3$. The other points on the
curve $f_\infty^=$ can be found using the same trick of modifying
the top coin $\G_r$. That is, let ${\G'}^{(n)}$ be as above but with
${\G'_r}^{(n)}=t$, where $t\in[0,1]$ is a parameter we can choose
freely. In the limit of $n\rightarrow \infty$ we find:
\be
{\A_r'}^{(\infty)}(t) &=& t \,\frac{1}{3} + (1-t),\\
{\B_r'}^{(\infty)}(t) &=& t \,\sqrt{\frac{1}{3}},\\
{\HH_r'}^{(\infty)}(t) &=& t \,\frac{1}{2}.
\ee
\noindent
The associated quantum weak coin-flipping parameters are
\be
P_A(t)&=& 1-\frac{t}{2},\label{eq:t1}\\
P_A^*(t)&=& \frac{2}{3}\, \frac{3-2t}{2-t},\label{eq:t2}\\
P_B^*(t)&=&\frac{2}{3}\, t.\label{eq:t3}
\ee

\noindent
These protocols correspond to the right half of the curve $f_\infty^=$
(i.e., the points $(z,f_\infty(z))$ for $z\in[1/3,1]$). The other half of
the curve can be obtained by symmetry between Alice and Bob. In the
Coin-Game formalism this symmetry arises by creating a new
$(n+1)$-Coin-Game, $\G'$, out of given $n$-Coin-Game, $\G$, by the rules
$\G'_r = 1$, $\G'_{0x}=\G'_{1x}=\G_x$ for $|x|<n$ and
$\G'_{0x}=\G'_{1x}=1-\G_x$ for $|x|=n$. In the language of protocols, we
are forcing Alice's first message to have no content, which is equivalent
to allowing Bob to begin the game.

The results can be best summarized by eliminating the variable $t$ from 
Eqs.~(\ref{eq:t1}--\ref{eq:t3}), which proves this section's main theorem:

\begin{theorem}
There exist quantum weak coin-flipping protocols that asymptotically
approach the curve
\be
P_A^* + P_B^* - \frac{3}{4} P_A^* P_B^* = 1
\ee
\noindent
in the limit of large number of messages. The corresponding probabilities of
winning when the game is played honestly are 
\be
P_A&=&\frac{3}{4}\, P_A^*\ \ \ \ \text{when $P_A^* \leq P_B^*$},\\
P_B&=&\frac{3}{4}\, P_B^*\ \ \ \ \text{when $P_A^* \geq P_B^*$}.
\ee
\end{theorem}

\subsection*{Implementing the optimal protocols}

Surprisingly, the optimal protocols identified above are significantly
easier to describe and implement than a generic protocol associated with a
random $n$-Coin-Game. Here we shall present a brief description of the
simplified protocol associated with the Coin-Games from
Eqs.~(\ref{eq:optg}--\ref{eq:optPB}).

We begin by fixing a security parameter $n$, which will lead to an $n+2$
message quantum protocol. For simplicity we assume that $n$ is even.

The first $n$ messages of the quantum protocol each involve one player
preparing a two qubit entangled state and sending one of the two qubits to
the other player. The two qubit states can be written as
\be
\sqrt{a_{i}} \ket{00} + \sqrt{1-a_{i}} \ket{11}
\ee
\noindent
for $i=1,\dots,n$, where
\be
a_i = 
\begin{cases}
\frac{n+1}{n+2}& i = 1,\cr
\frac{n-i+1}{n-i+3}& i \neq 1.
\end{cases}
\ee
\noindent
As usual Alice is in charge of sending the odd messages (and hence
preparing the odd numbered states) whereas Bob sends the even numbered
messages. 

At the end of the above procedure Alice and Bob should each have $n$
qubits, which can be expressed in a basis of $n$-bit strings with the most
significant bit corresponding to the first qubit sent or received.  They
now each perform a two-outcome measurement which can be described as
follows: let $S$ be the set of all $n$-bit strings such that the first
occurrence of the digit one, when the bits are examined from left to right,
appears at an even location, again counting from left to right (i.e., for
$n=4$ we have $S=\{0001, 0100, 0101, 0110, 0111\}$). The two outcome
measurement is given by the POVM elements
\be
E_0 = I - E_1,\qquad E_1 = \sum_{x\in S}\ket{x}\bra{x}.
\ee
As usual Alice wins on outcome zero and Bob wins on outcome one. Note that,
in essence, the first person to send a qubit in the ``one'' state is the winner
at this stage. However, the following cheat detection step will be
powerful enough to dissuade against the obvious cheating strategy.

Before outputting the final answer the party who won sends all their qubits
over to the losing party who then does an extra cheat-detecting two-outcome
measurement to verify that the $2n$ qubit state now in their possession is
the correct one (i.e., they project onto the state and its complement).
Unfortunately, this final step is likely to be very fairly difficult with
current technology for any $n>2$. 

In the end, the resulting protocol goes to a bias of $1/6$ as $n$ is taken to
infinity. For $n=4$ Bob's probability of winning by cheating is
$P_B^*=0.694$ whereas for $n=6$ we get $P_B^*=0.681$. Furthermore, Alice's
cheating is always restricted at $P_A^*=2/3$. Protocols with more symmetry
between Alice and Bob can also be described as above by changing the
coefficients $\{a_i\}$.

\section{Conclusions}

We have identified a large family of quantum protocols for weak coin
flipping, that are based on classical public-coin games. The family
contains protocols approaching the curve $P_A^* + P_B^* - \frac{3}{4} P_A^*
P_B^* = 1$, which can be reached asymptotically in the limit of large
number of messages. The most important of these protocols is symmetric
between Alice and Bob and achieves $P_A=P_B=1/2$ and $P_A^*=P_B^*=2/3$,
that is, it has a bias of $1/6$.

Furthermore, we have proven lower bounds for the bias achievable by
protocols in this family. In particular, $\max(P_A^*,P_B^*)\geq 2/3$ or
equivalently $\epsilon\geq 1/6$. These lower bounds show
that the protocols found above are optimal within their family.

Our lower bounds also establish a strict hierarchy among coin-flipping
protocols in our family with different number of messages. Admittedly, the
hierarchy is of little practical interest since a small number of messages
suffices in all cases to construct protocols that are reasonably close to
optimal.

Though the question of optimal bias for a general quantum weak
coin-flipping protocol remains open, we speculate that it might be possible
to show that every protocol is equivalent to one contained in the family
analyzed in this paper. Future work will be needed to verify this conjecture.

\begin{acknowledgments}

The author would like to thank Graeme Smith for his help in
proofreading this manuscript. This work was supported in part by the
National Science Foundation under grant number EIA-0086038 and by the
Department of Energy under grant number DE-FG03-92-ER40701.
 
\end{acknowledgments}


\appendix

\section{\label{sec:theorem}The Protocol}

The purpose of this appendix is to describe the $(n+1)$-message quantum weak
coin-flipping protocol associated to each $n$-Coin-Game. For each protocol
we shall also derive matching upper and lower bounds on the amount that
each party can cheat and thereby prove Theorem~\ref{theorem:proto}.

All the general ideas needed in this section have appeared previously in
Ref.~\cite{me2004}, though in a somewhat different notation. The new
elements of this appendix are:

\begin{enumerate}
\item Ref.~\cite{me2004} was restricted to $n$-Coin-Games where all the
binary nodes at the same depth had the same value (i.e, $\G_x=\G_{x'}$ if
$|x|=|x'|<n$). These variables were given the name $a_i$ so that
$\G_x=a_{|x|+1}$. In this section we lift the restriction and consider
general $n$-Coin-Games.
\item An upper bound on $P_A^*$ and $P_B^*$ was derived in
Ref.~\cite{me2004} but was not proven optimal. In this section we shall
derive a matching lower bound.
\end{enumerate}

Because most of the ideas here have been published elsewhere, we shall
simply prove the necessary facts in this section without providing the
intuition or motivation behind the constructions. For a more pedagogical
approach we refer the reader to Ref.~\cite{me2004}.

We begin by fixing an $n$-Coin-Game $\G$, which will be used throughout
this section. We also fix $\HH$, $\A$ and $\B$ as given by
Eqs.~(\ref{eq:h},\ref{eq:ab}). Because optimal protocols with $\HH_r=0$ and
$\HH_r=1$ are easy to construct even classically, for what follows we shall
assume that $0<\HH_r<1$.

To describe the quantum protocol associated with $\G$ we employ the
standard quantum communication model involving the Hilbert space
decomposition $\Hil_A\otimes\Hil_M\otimes\Hil_B$, where $\Hil_A$ is Alice's
private space, $\Hil_B$ is Bob's private space, and $\Hil_M$ is the space
used for passing messages. We further subdivide these spaces as follows:
\be
\Hil_A&=&\Hil_a\otimes\Hil_{a'}\otimes\Hil_{a c},\\
\Hil_B&=&\Hil_b\otimes\Hil_{b'}\otimes\Hil_{b c},\\
\Hil_M&=&\Hil_m\otimes\Hil_{m n}.
\ee
\noindent
The spaces $\Hil_a$ and $\Hil_b$ each consists of $n$ qubits and will be
used to store a binary string $x$ corresponding to a node in $\G$. The
individual qubits comprising each space will be referred to as $a_1$ through
$a_n$ and $b_1$ though $b_n$ respectively. The one-qubit space $\Hil_m$ will
be the primary means of communication between Alice and Bob, and will be
referred to as qubit $m$.

The rest of the spaces will only be used in the last pair of messages. The
spaces $\Hil_{a'}$, $\Hil_{b'}$ and $\Hil_{m n}$ each involve $n$ qubits
whereas $\Hil_{a c}$ and $\Hil_{b c}$ each contain one qubit.

Before describing the protocol we need to define a set of unitaries on
$\Hil_A\otimes\Hil_M$. We begin with the controlled rotations $R_{A,k}$
defined for $k=1,\dots,n$ by
\be
R_{A,k}=\sum_{\substack{x\cr|x|=k-1}}
\ket{x}\bra{x}_{a_1,\dots,a_{k-1}} \otimes U(\G_x)_{a_k,m},
\ee
\noindent
where
\be
U(z)=\mypmatrix{
\sqrt{z}   & 0           & 0           & -\sqrt{1-z}  \cr
0          & \sqrt{z}    & -\sqrt{1-z} & 0            \cr
0          & \sqrt{1-z}  & \sqrt{z}    & 0            \cr
\sqrt{1-z} & 0           & 0           & \sqrt{z}
}.
\label{eq:u}
\ee
\noindent
The subscripts on the operators and matrices indicate what qubits they act
on, and $R_{A,k}$ acts trivially on all qubits of $\Hil_{A}\otimes\Hil_{M}$
not explicitly mentioned. For the case $k=1$ the operator is not a
controlled rotation but rather a regular rotation using parameter $\G_r$.

We shall also need the controlled rotation
\beq
R_{A,E}=\sum_{\substack{x\cr|x|=n}}
\ket{x}\bra{x}_{a_1,\dots,a_n} \otimes 
\mypmatrix{
1-\G_x & -\G_x\cr
\G_x   & 1-\G_x}_{ac},\quad
\eeq
\noindent
which is unitary because $\G_x\in\{0,1\}$ for $|x|=n$. The gate is simply a
controlled-X applied to the qubit in space $\Hil_{ac}$, where the control
depends on a function of the qubits in $\Hil_a$. Note that $R_{A,E}$ can
also be defined as an operator acting purely on $\Hil_A$ rather than
$\Hil_{A}\otimes\Hil_{M}$.

Finally, define $S_{A,k}$ for $k=1,\dots,n$ to swap qubit $a_k$ with qubit
$m$:
\be
S_{A,k}=\text{SWAP}(a_k,m).
\ee
\noindent
We also need $T_{A,0}$ which swaps $\Hil_{a}$ with $\Hil_{m n}$
conditioned on qubit $a c$ being zero, and $T_{A,1}$ which swaps the space
$\Hil_{a'}$ with the space $\Hil_{n m}$ conditioned on qubit $a c$ being
one:
\be
T_{A,0} = \ket{0}\bra{0}_{a c} \otimes \text{SWAP}(\Hil_{a},\Hil_{m n}) 
+ \ket{1}\bra{1}_{a c} \otimes I,
\\
T_{A,1}= \ket{1}\bra{1}_{a c} \otimes \text{SWAP}(\Hil_{m n},\Hil_{a'}) 
+ \ket{0}\bra{0}_{a c} \otimes I.
\ee
\noindent
The first one is used to send the qubits in $\Hil_{a}$ when Alice
wins, whereas the second one is used to receive Bob's qubits and put them
in $\Hil_{a'}$ when Alice loses.

All the above operators act on Alice's Hilbert space. We can similarly
define the operators $R_{B,k}$, $R_{B,E}$, $S_{B,k}$ acting in the same way
on Bob's qubits. The operator $T_{B,0}$ however has to be defined to swap
$\Hil_{b'}$ with $\Hil_{m n}$ conditioned on qubit $b c$ being zero,
whereas $T_{B,1}$ swaps $\Hil_{b}$ with $\Hil_{m n}$ conditioned on qubit
$b c$ being one.

To characterize the final measurements it is useful to define the
probability tree $\PP$ by
\be
\PP_x=\begin{cases}
1&\text{if $x=r$},\cr
\G_y \PP_y&\text{if $x=y0$},\cr
(1-\G_y) \PP_y&\text{if $x=y1$}.
\end{cases}
\label{eq:p}
\ee
\noindent
That is, $\PP_x$ is the probability of reaching node $x$ when the classical
coin-flipping game associated with $\G$ is played honestly. We can now
define the two normalized states
\be
\ket{\psi_{A,1}} 
&=& \frac{1}{\sqrt{\HH_r}} \sum_{\substack{x\cr|x|=n\cr\G_x=1}} 
\sqrt{\PP_x} \ket{x}_{\Hil_a}\otimes\ket{x}_{\Hil_{a'}}
\otimes\ket{1}_{\Hil_{a c}},\nonumber\\
\ket{\psi_{B,0}} 
&=& \frac{1}{\sqrt{1-\HH_r}} \sum_{\substack{x\cr|x|=n\cr\G_x=0}} 
\sqrt{\PP_x} \ket{x}_{\Hil_b}\otimes\ket{x}_{\Hil_{b'}} 
\otimes\ket{0}_{\Hil_{b c}}.\nonumber\\
\ee

\noindent
The normalization is correct because $\HH_r$ is the probability of arriving
at a leaf $x$ such that $\G_x=1$, whereas $1-\HH_r$ is the probability of
arriving at a leaf with $\G_x=0$.
We are now ready to describe the main protocol.

\begin{proto1} Given an $n$-Coin-Game, $\G$, and the associated operators 
described above, define a quantum weak coin-flipping protocol by the
following steps:
\begin{enumerate}
\item Setup: Alice starts with $\Hil_A\otimes\Hil_M$ and Bob with $\Hil_B$.
They each initialize their space to the state $\ket{0}$.
\item First $n$ messages. For $k=1$ to $n$:
\begin{itemize}
\item If $k$ is odd, Alice applies $R_{A,k}$ and sends $\Hil_M$ to
Bob who applies $S_{B,k}$.
\item If $k$ is even, Bob applies $R_{B,k}$ and sends $\Hil_M$ to
Alice who applies $S_{A,k}$.
\end{itemize}
\item Alice applies $R_{A,E}$ to $\Hil_A$ and Bob applies $R_{B,E}$ 
to $\Hil_B$. No messages are needed for this step.
\item If Bob has $\Hil_M$ he sends it to Alice.
\item Alice applies $T_{A,0}$ and sends $\Hil_M$ to Bob who applies
$T_{B,0}$.
\item Bob applies $T_{B,1}$ and sends $\Hil_M$ to Alice who applies
$T_{A,1}$.
\item Alice measures using the two outcome POVM 
$\{I-\ket{\psi_{A,1}}\bra{\psi_{A,1}},\,\ket{\psi_{A,1}}\bra{\psi_{A,1}}\}$.
Bob measures the two outcome POVM 
$\{\ket{\psi_{B,0}}\bra{\psi_{B,0}},\,I-\ket{\psi_{B,0}}\bra{\psi_{B,0}}\}$.
They each output zero for the first outcome and one for the second.
\end{enumerate}
\end{proto1}

The basic intuition behind the protocol is that the first three steps above
is a quantum implementation of the classical public-coin coin-flipping
protocol associated with $\G$ described in Sec.~\ref{sec:not}. After $k$
messages the first $k$ bits of $\Hil_A$ contain a length $k$ string
indicating the depth $k$ node at which we are currently located. The
quantum amplitude associated with each such state is $\sqrt{\PP_x}$. Step 3
is a unitary realization of the measurement that looks at the $n$ bit
string $x$ corresponding to a leaf, and stores the classical coin outcome
in the qubit associated with  $\Hil_{a c}$ for Alice and $\Hil_{b c}$ for Bob.

The rest of the steps involve cheat detection. Effectively, the winner
declares victory immediately and then sends as much of their state as
possible to the other party. The losing party then checks that the state is
correct before accepting defeat.

Note that, as written, the above protocol takes either $n+2$ or $n+3$
messages. However, it is easy to see that the protocol can be run with only
$n+1$ messages. For starters, only the space $\Hil_{m}$ needs to be sent
back and forth in step 2, whereas only $\Hil_{m n}$ is used in steps 5 and
6. If we allow such a splitting, Alice starts with $\Hil_{m n}$ and step 4
is never needed. This reduces the protocol to $n+2$ messages always. But if
$n$ is odd then Alice ends up sending two messages in a row. The two
messages can be combined into a single longer message and therefore the
protocol only requires $n+1$ messages. We will also argue below that steps
5 and 6 can be interchanged, in which case when $n$ is even Bob sends two
messages in a row, and their merger leads again to a protocol with only
$n+1$ messages.

We turn to the task of describing the evolution of the game when both
players are honest. The action of $R_{A,k}$ entangles qubit $a_k$ with
qubit $m$, whereas $S_{B,k}$ swaps qubit $m$ with $b_k$. Their combined
effect is the transformation
\be
&&\sqrt{\PP_x} \ket{x}_{a_1,\dots,a_{k-1}} \otimes 
\ket{0}_{a_k}\otimes\ket{0}_{b_k}\\\nonumber
&&\longrightarrow 
\sqrt{\PP_{x0}} \ket{x}_{a_1,\dots,a_{k-1}} \otimes 
\ket{0}_{a_k}\otimes\ket{0}_{b_k}\\\nonumber
&&\qquad +
\sqrt{\PP_{x1}} \ket{x}_{a_1,\dots,a_{k-1}} \otimes
\ket{1}_{a_k}\otimes\ket{1}_{b_k}.
\ee

\noindent
The same effect occurs on even rounds when Alice's and Bob's actions are
reversed. Therefore, the state after the first $k$ passes through step 2 is
given by
\be
\ket{\psi_k}=\sum_{\substack{x\cr|x|=k}}&\sqrt{\PP_x}& 
\ket{x0\cdots0}_{\Hil_{a}}\otimes\ket{0}_{\Hil_{a'}\otimes\Hil_{a c}}
\\\nonumber
&&\otimes
\ket{x0\cdots0}_{\Hil_{b}}\otimes\ket{0}_{\Hil_{b'}\otimes\Hil_{b c}}
\otimes
\ket{0}_{\Hil_M},
\ee
\noindent
where there are $n-k$ zeroes following each $x$.

Step 3 simply has the effect of setting up the fair coin outcome in
$\Hil_{a c}$ and $\Hil_{b c}$:
\be
\ket{\psi_E}=\sum_{\substack{x\cr|x|=n}}&\sqrt{\PP_x}& 
\ket{x}_{\Hil_{a}}\otimes\ket{0}_{\Hil_{a'}}
\otimes\ket{\G_x}_{\Hil_{a c}}
\\\nonumber
&&\otimes
\ket{x}_{\Hil_{b}}\otimes\ket{0}_{\Hil_{b'}}
\otimes\ket{\G_x}_{\Hil_{b c}}
\otimes\ket{0}_{\Hil_M}.
\ee

Finally, when both players are honest, step 5 has the effect of moving
$\Hil_a$ to $\Hil_{b'}$ conditioned on qubits $a c$ and $b c$ both being
one.  Step 6 has the effect of swapping $\Hil_b$ to $\Hil_{a'}$ conditioned
on $a c$ and $b c$ being both zero. The final state of the protocol is
therefore:
\be
\ket{\psi_F}&\!\!\!\!=\!\!\!\!
&\sum_{\substack{x\cr|x|=n\cr\G_x=1}}\sqrt{\PP_x} 
\ket{x}_{\Hil_{a}}\otimes\ket{x}_{\Hil_{a'}}
\otimes\ket{1}_{\Hil_{a c}}
\\\nonumber
&&\qquad\otimes
\ket{0}_{\Hil_{b}}\otimes\ket{0}_{\Hil_{b'}}
\otimes\ket{1}_{\Hil_{b c}}
\otimes\ket{0}_{\Hil_M}
\\\nonumber
&&+\sum_{\substack{x\cr|x|=n\cr\G_x=0}}\sqrt{\PP_x} 
\ket{0}_{\Hil_{a}}\otimes\ket{0}_{\Hil_{a'}}
\otimes\ket{0}_{\Hil_{a c}}
\\\nonumber
&&\qquad\quad\otimes
\ket{x}_{\Hil_{b}}\otimes\ket{x}_{\Hil_{b'}}
\otimes\ket{0}_{\Hil_{b c}}
\otimes\ket{0}_{\Hil_M}
\\\nonumber
&\!\!\!\!=\!\!\!\!
& \sqrt{\HH_r} \ket{\psi_{A,1}}\otimes
\ket{0}_{\Hil_{b}\otimes\Hil_{b'}}
\otimes\ket{1}_{\Hil_{b c}}
\otimes\ket{0}_{\Hil_M}
\\\nonumber
&& + \sqrt{1-\HH_r} \ket{0}_{\Hil_{a}\otimes\Hil_{a'}}
\otimes\ket{0}_{\Hil_{a c}}
\otimes\ket{\psi_{B,0}}
\otimes\ket{0}_{\Hil_M}.
\ee

Because $\ket{\psi_{A,1}}$ is orthogonal to any state with the value zero
in register $\Hil_{a c}$ and $\ket{\psi_{B,1}}$ is orthogonal to any state
with the value one in register $\Hil_{b c}$, there are only two possible
outcomes for the final measurements:
\begin{itemize}
\item Alice obtains $I-\ket{\psi_{A,1}}\bra{\psi_{A,1}}$ and Bob obtains 
$\ket{\psi_{B,0}}\bra{\psi_{B,0}}$ in which case they both output
zero, that is, Alice wins. This happens with probability $1-\HH_r$.
\item Alice obtains $\ket{\psi_{A,1}}\bra{\psi_{A,1}}$ and Bob obtains 
$I-\ket{\psi_{B,0}}\bra{\psi_{B,0}}$ in which case they both output
one, that is, Bob wins. This happens with probability $\HH_r$.
\end{itemize}

\noindent
We have therefore proven the following lemma:

\begin{lemma}
When playing Protocol 1 honestly, Alice's and Bob's outputs are perfectly
correlated and satisfy
\be
P_A=1-\HH_r,\qquad P_B=\HH_r.
\ee
\end{lemma}

\subsection{Reformulation as an SDP}

We now turn to the analysis of the advantage that a cheating player can
attain. Specifically, we shall focus on the case of honest Alice and
cheating Bob. The case where Alice is cheating is fairly similar and will
be derived at the end of the appendix from the case of cheating Bob.

When Bob is cheating we don't know exactly what operations (unitaries,
measurements, or superoperators) he may be applying to his qubits. In fact,
we don't even know how many qubits he may have in his laboratory. We shall
therefore focus only on the evolution of the qubits under Alice's
control. This approach, first advocated by Kitaev
\cite{Kitaev}, will transform the maximization over Bob's cheating
strategies into a semidefinite program (SDP).

Let $\rho_0$ be the initial state of all qubits under Alice's control,
that is, it is a density operator on $\Hil_A\otimes\Hil_M$. Let
$\rho_1,\dots,\rho_n$ be the state of the qubits under Alice's control
after each of the $n$ passes through step 2. Note that $\rho_k$ is a
density operator for $\Hil_A$ when $k$ is odd, and for
$\Hil_A\otimes\Hil_M$ when $k$ is even. Finally let $\rho_E$ be the state
of $\Hil_A\otimes\Hil_M$ at the end of step 4 and let $\rho_F$ be the state
of $\Hil_A\otimes\Hil_M$ at the end of step 6.

Because Alice initializes her own qubits as prescribed by the protocol
without interference from Bob, their initial state is given by
\be
\rho_0 = \ket{0}\bra{0}_{\Hil_A\otimes\Hil_M}.
\label{eq:r0}
\ee

For odd $k$, Alice first applies the unitary $R_{A,k}$ and then sends
$\Hil_M$ to Bob, leaving the state
\beq
\rho_k = \Tr_M\lp[R_{A,k}\,\rho_{k-1} R_{A,k}^{-1}\rp]\qquad 
\text{(for $k$ odd)}.
\label{eq:rodd}
\eeq
\noindent
For even $k$, we can't fully characterize $\rho_k$ in terms of $\rho_{k-1}$
but we know that given $\rho_k$, if we undo the swap $S_{A,k}$ and then
send back $\Hil_M$ we must end up with $\rho_{k-1}$, therefore
\beq
\Tr_M\lp[S_{A,k}^{-1}\,\rho_k S_{A,k}\rp]=\rho_{k-1}\qquad 
\text{(for $k$ even)}.
\label{eq:reven}
\eeq

Step 3 only involved the use of $R_{A,E}$, a unitary on $\Hil_A$.
Step 4, the recovery of $\Hil_M$, is only needed when $n$ is
odd. Therefore,
\be
\rho_E &=& R_{A,E}\, \rho_n R_{A,E}^{-1}\qquad \text{for $n$ even},\\
\Tr_M \rho_E &=& R_{A,E}\, \rho_n R_{A,E}^{-1}\qquad \text{for $n$ odd}.
\ee

Finally, the state of the qubits on $\Hil_A$ after applying $T_{A,0}$ to
$\rho_E$ must equal the state $\rho_F$ if we undo $T_{A,1}$ (because as
usual, Bob has no effect on Alice's qubits):
\be
\Tr_M\lp[T_{A,1}^{-1}\, \rho_F T_{A,1}\rp]=
\Tr_M\lp[T_{A,0}\, \rho_E T_{A,0}^{-1}\rp].
\label{eq:rf}
\ee

The probability that Bob wins is given by the final measurement
\be
\Tr\lp[\ket{\psi_{A,1}}\bra{\psi_{A,1}}\,\rho_F\rp],
\label{eq:max}
\ee
\noindent
where it is understood that $\ket{\psi_{A,1}}\bra{\psi_{A,1}}$ can be
extended to an operator on $\Hil_A\otimes\Hil_M$ by tensoring with the
identity $I_M$. 

The preceding arguments show that no matter what cheating strategy Bob
employs, the sequence of states for Alice's qubits must satisfy the above
equations, and therefore $P_B^*$ is upper bounded by the maximum of
Eq.~(\ref{eq:max}) over all assignments to the variables
$\rho_0,\dots,\rho_n$, $\rho_E$, $\rho_F$ consistent with the above
equations. It is also not hard to see that Bob can achieve any set of
density matrices consistent with the above equations by maintaining the
purification of Alice's state. As this reduction from maximization over
cheating strategies to SDP has already appeared in the literature
\cite{Kitaev,Ambainis2003,me2004} we won't belabor the point and simply
state the lemma we have proven:

\begin{lemma}
The maximum probability with which Bob can win by cheating in Protocol 1 is
given by the solution of the SDP:
\be
P_B^*= \max \Tr\lp[\ket{\psi_{A,1}}\bra{\psi_{A,1}}\,\rho_F\rp],
\ee
\noindent
over the positive semidefinite variables
$\rho_0,\dots,\rho_n$, $\rho_E$, $\rho_F$ subject to the constraints of
Eqs.~(\ref{eq:r0}--\ref{eq:rf}).
\end{lemma}

The security of the above result depends solely on the laws of quantum
mechanics and the assumption that Bob cannot directly influence the qubits
in Alice's laboratory. We note that we are assuming, as is usual in
coin-flipping protocols, that Alice can measure the size of the Hilbert
space $\Hil_M$ (i.e., the number of qubits sent by Bob in each message) and
that if at any point she receives more or less than the required number of
qubits she aborts the protocol and declares herself the winner. The optimal
strategy for Bob involves sending the right number of qubits in each
message and therefore is described by the above formalism.

It will be important below to know that we can exchange steps 5 and 6.
This would work as follows: given $\rho_E$ we send $\Hil_M$ to Bob, who is
supposed to apply $T_{B,1}$ to his qubits. Upon return, Alice applies 
$T_{A,1}$ followed by $T_{A,0}$ ending up with state $\rho_F'$ satisfying
\be
\Tr_M\lp[T_{A,1}^{-1}T_{A,0}^{-1}\,\rho_F' T_{A,0} T_{A,1}\rp]=
\Tr_M\lp[\rho_E\rp].
\label{eq:rfp}
\ee
\noindent
The final measurement can be done immediately before sending $\Hil_M$ to
Bob because it only has support on $\Hil_A$. The probability of Bob winning
is
\be
\Tr\lp[\ket{\psi_{A,1}}\bra{\psi_{A,1}}\,\rho_F'\rp].
\ee
\noindent
However, $\ket{\psi_{A,1}}$ only has support on the space where qubit $a c$
is one, and in this subspace $T_{A,0}$ acts trivially (and $T_{A,0}$ and
$T_{A,1}$ commute). Applying projectors to both sides of the
Eq.~(\ref{eq:rf}) and Eq.~(\ref{eq:rfp}) we see that both SDPs are
equivalent, and therefore steps 5 and 6 are interchangeable, at least from
the perspective of honest Alice.

\subsection{Lower Bounds}

To find a lower bound on $P_B^*$ we shall describe a specific assignment of
the variables $\rho$ that satisfies the above equations, and from it
calculate $\Tr\lp[\ket{\psi_{A,1}}\bra{\psi_{A,1}}\,\rho_F\rp]$. Because
$P_B^*$ is a maximum over such assignments, this will serve as a lower bound. 

Let
\beq
\rho_k = \begin{cases}
\sigma_k\otimes \ket{0}\bra{0}_{a_{k+1},\dots,a_{n}} \otimes 
\ket{0}\bra{0}_{\Hil_{a'}\otimes\Hil_{a c}}&\text{$k$ odd}\cr
\sigma_k\otimes \ket{0}\bra{0}_{a_{k+1},\dots,a_{n}} \otimes 
\ket{0}\bra{0}_{\Hil_{a'}\otimes\Hil_{a c}\otimes\Hil_M}&\text{$k$ even}
\end{cases}
\eeq
\noindent
where $\sigma_k$ is a density operator for qubits $a_1$ through
$a_k$. The operators $\rho_1$ through $\rho_n$ satisfy
Eqs.~(\ref{eq:rodd},\ref{eq:reven}) provided that
\beq
\sigma_k = \Tr_m\lp[ R_{A,k} \lp(\sigma_{k-1} \otimes
\ket{0}\bra{0}_{a_k,m} \rp) 
R_{A,k}^{-1}\rp]\quad \text{(for $k$ odd)}
\label{eq:sigodd}
\eeq
\noindent
where $\sigma_0=1$ is the unit, and
\beq
\Tr_{a_k}\lp[\sigma_k \rp]=\sigma_{k-1}\qquad\qquad \text{(for $k$ even)}.
\label{eq:sigeven}
\eeq

\noindent
The $\sigma$ operators above will be defined using a tree variable $\W$
given by the equation
\beq
\W_x=\begin{cases}
1&x=r\cr
\G_y W_y&\text{$x=y0$ and $|x|$ odd}\cr 
(1-\G_y) W_y&\text{$x=y1$ and $|x|$ odd}\cr 
\G_y \B_x^2 W_y/ \B_y &\text{$x=y0$ and $|x|$ even}\cr 
(1-\G_y) \B_x^2 W_y/ \B_y &\text{$x=y1$ and $|x|$ even}
\end{cases}
\eeq
\noindent
which is based on the weight matrix $W$ of Ref.~\cite{me2004}. Note that,
though it is possible for $\B_y$ to be zero, this can only occur if both
$\B_{y0}$ and $\B_{y1}$ are zero as well, in this case we define
$\W_{y0}=\W_{y1}=0$, which resolves the potential division by zero.

Because $\B$ is computed bottom-up, whereas $\W$ is computed top-down,
every node of $\W$ depends on the complete $n$-Coin-Game assignment
$\G$. The appearance at every node of such global information about the
protocol is crucial for optimal solutions of these SDPs and will also occur
with the tree variable $\Z$ defined below in the section on upper bounds.

Define the $\sigma$ operators as diagonal matrices with entries given by
\be
\bra{x}\sigma_k\ket{x}=\W_x\qquad\text{for $|x|=k$}.
\ee
\noindent
The requirements of Eq.~(\ref{eq:sigodd}) are satisfied if
\beq
\W_{y0} = \G_{y}\W_{y}\quad\text{and}\quad\W_{y1} = (1-\G_{y})\W_{y}
\quad\text{(for $|y|$ even),}
\eeq
whereas Eq.~(\ref{eq:sigeven}) only imposes the weaker requirement
\be
\W_y = \W_{y0}+\W_{y1} \quad\text{(for $|y|$ odd),}
\ee
\noindent
both of which are clearly satisfied by $\W$. We have therefore outlined a
valid cheating strategy for Bob through step 2.

The next two steps will follow the protocol exactly, in which case
the operator $\rho_E$ follows from $\rho_n$ by adjusting the space 
$\Hil_{a c}$:
\be
\rho_E &=& \sum_{x} \W_x \ket{x}\bra{x}_{\Hil_a} \otimes 
\ket{0}\bra{0}_{\Hil_{a'}} \otimes \ket{\G_x}\bra{\G_x}_{\Hil_{a c}}
\nonumber\\
&&\otimes \ket{0}\bra{0}_{\Hil_{M}}.
\ee

Finally, in the last steps, conditioned on qubit $a c$ being zero Alice
sends her state to Bob. Conditioned on qubit $a c$ being one, Bob returns
the purification of the remaining qubits, so the final state is:
\be
\rho_F&=&\ket{\phi_1}\bra{\phi_1}_{\Hil_a\otimes\Hil_{a'}}\otimes
\ket{1}\bra{1}_{\Hil_{a c}}\otimes \ket{0}\bra{0}_{\Hil_M}
\\\nonumber
&&+ C_0
\ket{0}\bra{0}_{\Hil_a\otimes\Hil_{a'}}\otimes
\ket{0}\bra{0}_{\Hil_{a c}}\otimes \ket{0}\bra{0}_{\Hil_M},
\ee
\noindent
where $C_0$ is an unimportant constant (equal to the sum of $\W_x$ for all
$x$ such that $\G_x=0$), and $\ket{\phi_1}$ is the unnormalized state given
by
\be
\ket{\phi_1}=\sum_{\substack{x\cr|x|=n\cr\G_x=1}}\sqrt{\W_x} 
\ket{x}_{\Hil_a}\otimes\ket{x}_{\Hil_{a'}}.
\ee

Bob's probability of winning is given by 
\be
p&=&\lp| \lp(\bra{\phi_1}_{\Hil_a\otimes\Hil_{a'}}\otimes\bra{1}_{\Hil_{ac}}
\rp)\ket{\psi_{A,1}} \rp|^2
\\\nonumber
&=& \lp| \sum_{\substack{x\cr|x|=n}} \G_x \sqrt{\W_x \PP_x} \rp|^2/\HH_r,
\ee
\noindent
where the factor of $\G_x$ ensures that the sum is only taken over strings
$x$ satisfying $\G_x=1$.

While the expression for computing $p$ seems rather daunting, we shall show
in a moment that when properly written, it is a conserved quantity that has
the same value at every depth in the tree. We begin with the following two
observations: for $|y|$ even
\be
&&\sqrt{\B_{y0}}  \sqrt{\W_{y0} \PP_{y0}} + 
\sqrt{\B_{y1}}  \sqrt{\W_{y1} \PP_{y1}}
\nonumber\\
&&= \lp( \G_y \sqrt{\B_{y0}} + (1-\G_y) \sqrt{\B_{y1}} \rp)\sqrt{\W_y \PP_y}
\nonumber\\
&&= \B_y \sqrt{\W_y \PP_y}
\ee
\noindent
whereas for $|y|$ odd we have
\be
&&\B_{y0} \sqrt{\W_{y0} \PP_{y0}} + \B_{y1}  \sqrt{\W_{y1} \PP_{y1}}
\nonumber\\
&& = \lp( \G_y \B_{y0}^2 + (1-\G_y) \B_{y1}^2\rp) \sqrt{\W_y \PP_y/\B_y}
\nonumber\\
&& = \sqrt{\B_y} \sqrt{\W_y \PP_y}.
\ee
\noindent
For the special case when $\B_y=0$ the equation is also valid as it reads
$0+0=0$. By induction, we can obtain the following result
\beq
\B_r \sqrt{\W_r \PP_r} =\begin{cases}
\sum_{x;|x|=k} \B_x \sqrt{\W_x \PP_x}&
\text{for any even $k$,}\cr
\sum_{x;|x|=k} \sqrt{\B_x} \sqrt{\W_x \PP_x}&
\text{for any odd $k$,}
\end{cases}
\label{eq:cons1}
\eeq
\noindent
where as usual $0\leq k\leq n$. In particular, because for $|x|=n$ we have
$\G_x=\B_x=\sqrt{\B_x}\in\{0,1\}$ we have shown that 
$p=\lp|\B_r \sqrt{\W_r \PP_r}\rp|^2/\HH_r$, which is the probability with
which Bob can win the coin-flip by cheating using the strategy outlined
above. Since $\W_r=\PP_r=1$ we have proven the desired lower bound:
\begin{lemma}
For Protocol 1:
\be
P_B^* \geq \frac{\B_r^2}{\HH_r}.
\ee 
\end{lemma}

\subsection{Upper Bounds}

We shall prove an upper bound by exhibiting a solution to the dual SDP.
We use the derivation of the dual in Ref.~\cite{Ambainis2003}, though
a direct derivation (as was done in Ref.~\cite{me2004}) would be fairly
simple as well. 

Our protocol can be rewritten in the notation of Ref.~\cite{Ambainis2003}.
Let $m=\lfloor (n+1)/2\rfloor$ and define $U_{A,1}=R_{A,1}$,
$U_{A,j}=R_{A,2j-1}S_{A,2j-2}$ for $j=2,\dots,m$, $U_{A,m+1}= T_{A,0}
R_{A,E} S_{A,n}$ (or if $n$ is odd just $U_{A,m+1}= T_{A,0} R_{A,E}$) and
$U_{A,m+2}=T_{A,1}$. The final measurement is
$\Pi_{A,1}=\ket{\psi_{A,1}}\bra{\psi_{A,1}}$. In this notation, we are
looking for the maximum of $\Tr[\Pi_{A,1}\,\rho_{A,m+2}]$ over assignments of
the positive semidefinite variables $\rho_{A,0},\dots,\rho_{A,m+2}$
satisfying:
\be
\Tr_{M}\lp[ \rho_{A,j} \rp] = 
\Tr_{M}\lp[U_{A,j}\, \rho_{A,j-1}\, U_{A,j}^{-1} \rp]
\ee
\noindent
for $j=1,\dots,m+2$ and $\Tr_{M}
[\rho_{A,0}]=\ket{0}\bra{0}_{\Hil_A}$. The initial condition for
$\rho_{A,0}$ (rather than the usual
$\rho_{A,0}=\ket{0}\bra{0}_{\Hil_A\otimes\Hil_M}$) simply gives Bob a little
more cheating power (i.e., to initialize $\Hil_M$) but this is acceptable
as we are now focusing on deriving upper bounds on $P_B^*$ and this extra
cheating power will not be helpful.

The dual SDP is given by  Lemma 11 of Ref.~\cite{Ambainis2003} as the
minimization of $\bra{0} Y_{A,0} \ket{0}$, subject to
\beq
Y_{A,j}\otimes I_{\Hil_M}\geq U_{A,j+1}^{-1}\lp( Y_{A,j+1}\otimes
I_{\Hil_M}\rp)U_{A,j+1} 
\eeq
\noindent
for $0\leq j \leq m+1$, where $Y_0,\dots,Y_{m+1}$ are Hermitian operators
on $\Hil_A$ and $Y_{A,m+2}\equiv\Pi_{A,1}$. Because this is the dual SDP to
the original coin-flipping SDP corresponding to Protocol 1, any assignment
of the variables $Y_{A,i}$ that satisfies the constraints will produce a
value of $\bra{0} Y_{A,0} \ket{0}$ that is an upper bound on
$P_B^*$. However, rather than finding a solution to the above dual SDP, we
shall study a modified, but equivalent, SDP:

\begin{lemma}
Let $Z_0,\dots,Z_{n+2}$ be a set of Hermitian matrices, defined on $\Hil_A$,
satisfying the following equations:
\be
Z_k\otimes I_{\Hil_M}\geq& R_{A,k+1}^{-1} \lp(Z_{k+1} \otimes I_{\Hil_M}\rp)
R_{A,k+1} \quad &\text{($k$ even),}
\nonumber\\
Z_k\otimes I_{\Hil_M}\geq& S_{A,k+1}^{-1} \lp(Z_{k+1} \otimes I_{\Hil_M}\rp)
S_{A,k+1} \quad &\text{($k$ odd),}
\nonumber\\
\label{eq:zevenodd}
\ee
\noindent
where $0\leq k<n$, and
\be
Z_{n}\otimes I_{\Hil_M}&\geq& R_{A,E}^{-1} \lp(Z_{n+1} \otimes I_{\Hil_M}\rp)
R_{A,E},
\label{eq:zn}
\\
Z_{n+1}\otimes I_{\Hil_M}&\geq& T_{A,0}^{-1} \lp(Z_{n+2} 
\otimes I_{\Hil_M}\rp) T_{A,0},
\label{eq:zn1}
\\
Z_{n+2}\otimes I_{\Hil_M}&\geq& T_{A,1}^{-1} 
\lp(\ket{\psi_{A,1}}\bra{\psi_{A,1}} \otimes I_{\Hil_M}\rp)
T_{A,1}.\qquad
\label{eq:zn2}
\ee
then $\beta\equiv\bra{0}Z_0\ket{0}$ is an upper bound on $P_B^*$.
\end{lemma}

The proof follows by noting that given a set of $Z_0,\dots,Z_{n+2}$
satisfying the above equations, we can set $Y_0=Z_0$, $Y_j=Z_{2j-1}$ for
$j=1,\dots,m$ and $Y_{m+1}=Z_{n+2}$ to obtain a solution with the same
minimum as the original dual SDP.

We introduce a new variable, defined on a tree of depth $n$, which shall be
used in constructing solutions of the dual SDP:
\be
\Z_x=\begin{cases}
\B_r^2/\HH_r&x=r\cr
\sqrt{\B_x} Z_y/\B_y&\text{$|x|$ odd}\cr 
Z_y &\text{$|x|$ even}
\end{cases}
\ee
\noindent
where $y$ is the parent node of $x$ (i.e., either $x=y0$ or $x=y1$). Once
again we resolve the division by zero by declaring $\Z_{y0}=\Z_{y1}=0$
whenever $\B_y=0$ and $|y|$ is even.

We begin the description of the solution to the dual SDP by choosing
\be
Z_{n+2} = \sum_{\substack{x\cr|x|=n\cr\G_x=1}} 
\Z_{x} \ket{x}\bra{x}_{\Hil_{a}}
\otimes I_{\Hil_{a'}}\otimes \ket{1}\bra{1}_{\Hil_{a c}}.
\ee
To verify that $Z_{n+2}$ satisfies Eq.~(\ref{eq:zn2}), we note that we can
move the unitary operators $T_{A,1}$ to the left hand side of the equation,
where they act trivially (i.e., they exchange $I_{\Hil_{a'}}$ with
$I_{\Hil_{m n}}$). We are left with the task of proving $Z_{n+2}\geq
\ket{\psi_{A,1}}\bra{\psi_{A,1}}$.

It is sufficient to show that
\be
Z_{n+2} + \epsilon I_{\Hil_A} \geq \ket{\psi_{A,1}}\bra{\psi_{A,1}}
\ee
for every $\epsilon>0$. Because $Z_{n+2}$ is non-negative, the
left-hand-side above is positive definite. We can rescale our space by
$(Z_{n+2} + \epsilon I_{\Hil_A})^{-1/2}$ to obtain the equivalent equation
\beq
I \geq \lp(Z_{n+2} + \epsilon I_{\Hil_A}\rp)^{-\frac{1}{2}}
 \ket{\psi_{A,1}}\bra{\psi_{A,1}} \lp(Z_{n+2} + 
\epsilon I_{\Hil_A}\rp)^{-\frac{1}{2}}.
\eeq
\noindent
The right-hand-side of the above equation has only one non-zero eigenvalue,
it is therefore sufficient to check that
\be
1 \geq \bra{\psi_{A,1}} \lp(Z_{n+2} + \epsilon I_{\Hil_A}\rp)^{-1} 
\ket{\psi_{A,1}}.
\ee
\noindent
We need to study the quantity
\beq
\bra{\psi_{A,1}} \lp(Z_{n+2} + \epsilon I_{\Hil_A}\rp)^{-1} 
\ket{\psi_{A,1}}
=  \sum_{\substack{x\cr|x|=n\cr\G_x=1}} \frac{\PP_x}{\HH_r (\Z_x+\epsilon)}
\eeq
\noindent
which once again is related to a conserved quantity at every level of the
tree. However, we first note the following properties which can be checked
directly from the definitions:
\begin{itemize}
\item $\PP_x>0$ implies $\PP_y>0$ for every node $y$ that has $x$ as a
descendant.
\item $\PP_x>0$ and $\B_x>0$ implies that $\B_y>0$ for every
node $y$ that has $x$ as a descendant.
\item $\PP_x>0$ and $\B_x>0$ implies $\Z_x>0$.
\end{itemize}

We can now remove $\epsilon$ from the above expression, because if $\Z_x=0$
then either $\PP_x=0$ or $\B_x=0$ (which implies $\G_x=0$):
\beq
\bra{\psi_{A,1}} \lp(Z_{n+2} + \epsilon I_{\Hil_A}\rp)^{-1} 
\ket{\psi_{A,1}}
\leq  \sum_{\substack{x\cr|x|=n\cr\Z_x>1}} \frac{\G_x \PP_x}{\HH_r \Z_x}
\eeq
\noindent
where the factor $\G_x$ imposes the condition $\G_x=1$, and the condition 
$\Z_x>0$ has been moved into the sum.

If $|y|$ is odd and $\Z_y>0$ we have
\be
\frac{\B_{y0}^2 \PP_{y0}}{\Z_{y0}} + \frac{\B_{y1}^2 \PP_{y1}}{\Z_{y1}} 
&=&\lp(\B_{y0}^2 \G_{y}  + \B_{y1}^2 (1-\G_{y}) \rp)
\frac{\PP_{y}}{\Z_{y}}
\nonumber\\
&=&\frac{\B_{y} \PP_{y}}{\Z_{y}},
\ee
\noindent
where the left-hand side is well defined because $\Z_{y0}=\Z_{y1}=\Z_y>0$.
If $|y|$ is even we have
\be
\frac{\B_{y0} \PP_{y0}}{\Z_{y0}} + \frac{\B_{y1} \PP_{y1}}{\Z_{y1}} 
\!&=&\!\lp(\sqrt{\B_{y0}} \G_{y}  + \sqrt{\B_{y1}} (1-\G_{y}) \rp)
\frac{\B_y \PP_{y}}{\Z_{y}}
\nonumber\\
&=&\frac{\B_{y}^2 \PP_{y}}{\Z_{y}}.
\ee
\noindent
Even if $\Z_y>0$ it is possible for either $\Z_{y0}$ or $\Z_{y1}$ (or both)
to be zero. If both are zero, then so is $\B_y \PP_y$. If only one of them
is zero (say $\Z_{y0}$) then the equation is still valid with the offending
term removed (that is, $\B_{y1} \PP_{y1}/\Z_{y1} = \B_{y}^2
\PP_{y}/\Z_{y}$). Using induction, we can prove
\beq
1 = \frac{\B_r^2 \PP_r}{\HH_r \Z_r} = \begin{cases}
\sum_{x;|x|=k;\Z_x>0} \frac{\B_x^2 \PP_x}{\HH_r \Z_x}&
\text{for any even k}\cr
\sum_{x;|x|=k;\Z_x>0} \frac{\B_x \PP_x}{\HH_r \Z_x}&
\text{for any odd $k$}
\end{cases}
\label{eq:cons2}
\eeq
\noindent
and in particular, because $\G_x=\B_x=\B_x^2$ for $|x|=n$ we have shown
$\bra{\psi_{A,1}} \lp(Z_{n+2} + \epsilon I_{\Hil_A}\rp)^{-1} 
\ket{\psi_{A,1}}\leq 1$ for every $\epsilon>0$, thus completing the proof that
our choice for $Z_{n+2}$ satisfies the requirement imposed by
Eq.~(\ref{eq:zn2}).

The next few requirements are easier to check. Since $Z_{n+2}$ only
has support on the space in which qubit $a c$ is one, on which $T_{A,0}$
acts trivially, we can satisfy Eq.~(\ref{eq:zn1}) by choosing
\be
Z_{n+1} &=& \sum_{\substack{x\cr|x|=n}} 
\Z_{x} \ket{x}\bra{x}_{\Hil_{a}}
\otimes I_{\Hil_{a'}}\otimes \ket{\G_x}\bra{\G_x}_{\Hil_{a c}}
\nonumber\\
&\geq& Z_{n+2},
\ee
\noindent
where the inequality follows because we have simply included the
(non-negative) coefficients for the states with $\G_x=0$.

The unitary $R_{A,E}$ operates only on the space $\Hil_{A}$ hence
 Eq.~(\ref{eq:zn}) can be satisfied by choosing
\be
Z_{n} &=& R_{A,E}^{-1} Z_{n+1} R_{A,E}
\\\nonumber
&=& \sum_{\substack{x\cr|x|=n}} 
\Z_{x} \ket{x}\bra{x}_{\Hil_{a}}
\otimes I_{\Hil_{a'}}\otimes \ket{0}\bra{0}_{\Hil_{a c}}.
\ee

Finally, fix a new parameter $\epsilon'>0$, and define
\be
Z_k &=& \sum_{\substack{x\cr|x|=k}} \lp( \Z_{x} + \frac{(n-k)\epsilon'}{n}
\rp) \ket{x}\bra{x}_{a_1,\dots,a_k}
\\\nonumber
&&\qquad\otimes \ket{0}\bra{0}_{\Hil_{a_{k+1},\dots,a_n}
\otimes\Hil_{a'}\otimes \Hil_{a c}}
\\\nonumber
&&+ C_k\, I_{a_1,\dots,a_k}\otimes \lp( I - \ket{0}\bra{0}\rp)_{
\Hil_{a_{k+1},\dots,a_n}\otimes\Hil_{a'}\otimes \Hil_{a c}},
\ee
\noindent
for $k=0,\dots,n-1$. The constants $C_k$ will be defined recursively below,
starting with $C_{n-1}$. For $k=0$ the above should be interpreted as
\be
Z_0 = \lp( \Z_r + \epsilon'\rp) \ket{0}\bra{0}_{\Hil_A} 
+ C_0 \lp(I- \ket{0}\bra{0}\rp)_{\Hil_A}.
\ee

In order to prove that our solution to the dual SDP is valid, all that
remains is to check Eq.~(\ref{eq:zevenodd}). The case of $k$ odd is fairly
simple because $\Z_y=\Z_{y0}=\Z_{y1}$ for $|y|$ odd, therefore qubit
$a_{k+1}$ of $Z_{k+1}$ is unentangled with the rest of the qubits and its
state is the identity density matrix (i.e., $Z_{k+1}=I_{a_{k+1}} \otimes
Z'$ where $Z'$ is an operator on the rest of the qubits). As the swap
operator $S_{A,k+1}$ acts trivially on $Z_{k+1}\otimes I_{\Hil_M}$, it is
sufficient to check $Z_{k}\geq Z_{k+1}$, which is satisfied if $C_{k}\geq
C_{k+1}$. For the special case of $k=n-1$ (and $n$ even) it suffices to
choose $C_k\geq\max \Z_x$ where the maximum is taken over all strings $x$ such
that $|x|=n$.

What remains to be proven is Eq.~(\ref{eq:zevenodd}) for the case of even
$k$. Fix some even value of $k$ and let $\alpha = Z_k\otimes I_{\Hil_M}$
and $\beta = R_{A,k+1}^{-1} \lp(Z_{k+1} \otimes I_{\Hil_M}\rp) R_{A,k+1}$.
There are just the left- and right-hand sides of the equation we are trying
to prove: $\alpha\geq \beta$. Define the projector
\beq
\Pi = I_{a_1,\dots,a_{k}}\otimes
\ket{0}\bra{0}_{\Hil_{a_{k+1},\dots,a_n}\otimes\Hil_a'
\otimes\Hil_{a c}}\otimes I_{\Hil_M}.
\eeq
\noindent
We shall prove in a moment $\Pi (\alpha - \beta) \Pi =
\frac{\epsilon'}{n}\Pi$. It is also easy to see that $\Pi \alpha
(I-\Pi)=(I-\Pi)\alpha \Pi = 0$ and $(I-\Pi)\alpha (I- \Pi) = C_k (I- \Pi)$.
Under these conditions, it is always possible to choose a large enough
$C_k$ so that $\alpha\geq\beta$, which defines $C_k$ in terms of $C_{k+1}$
(except for $C_{n-1}$ which can be defined directly from $Z_n$).  For a
proof, see for instance the proof of Lemma 3 in Ref.~\cite{me2004}.

To prove $\Pi (\alpha - \beta) \Pi = \frac{\epsilon'}{n}\Pi$ we need to
study the effect of the unitary $R_{A,k+1}$ on $Z_{n+1}$. The expression
has the form of a sum of $\ket{x}\bra{x}_{a_1,\dots,a_k}$ tensored with
\beq
U(\G_x)^{-1} \lp[ \lp( \Z_{x0} \ket{0}\bra{0}_{a_{k+1}} + 
\Z_{x1} \ket{1}\bra{1}_{a_{k+1}} \rp) \otimes I_m\rp] U(\G_x)
\eeq
\noindent
for $|x|=k$, where $U(z)$ is defined by Eq.~(\ref{eq:u}). The component of the
above that survives the projection $\Pi$ has the form
\be
&&\lp( \G_x \Z_{x0} + (1-\G_x) \Z_{x1} \rp) \ket{0}\bra{0}_{a_{k+1}}\otimes I_m
\nonumber\\
&&= \lp( \G_x \sqrt{\B_{x0}} + (1-\G_x) \sqrt{\B_{x1}} \rp)
\frac{\Z_x}{\B_x} \ket{0}\bra{0}_{a_{k+1}}\otimes I_m
\nonumber\\
&&= \Z_x \ket{0}\bra{0}_{a_{k+1}}\otimes I_m.
\ee
\noindent
It is now straightforward to check that $\Pi\alpha\Pi = \Pi \beta \Pi +
\frac{\epsilon'}{n}\Pi$, completing the proof that our choice of $Z_k$
satisfies Eq.~(\ref{eq:zevenodd}).

Note that, while the original protocol only depends on the first column of
the matrix $U(z)$, the above calculation involved the entire matrix. The
reason for this is that when transforming from the SDP involving the $Y$
variables to the SDP involving the $Z$ variables we gave Bob a small amount
of extra cheating power to set the qubits in $\Hil_M$ between application
of $S_{A,k}$ and $R_{A,k+1}$, in which case the full matrix $U(z)$ becomes
important. However, since the upper bound derived in this section matches
the lower bound from the last section, it should be clear that such extra
power is not useful.

The result thus far is the description of a set of variables
$Z_0,\dots,Z_{n+2}$ satisfying the equations of the dual SDP. This gives us
an upper bound $P_B^* \leq \beta=\bra{0}Z_0\ket{0} = Z_r +
\epsilon'$. However, since $\epsilon'>0$ is arbitrary, we have proven

\begin{lemma}
For Protocol 1:
\be
P_B^*\leq \frac{\B_r^2}{\HH_r}.
\ee
\end{lemma}

\subsection{Honest Bob v Cheating Alice}

The analysis of the case of honest Bob and cheating Alice is fairly similar
to the above calculations. Fortunately, we can exploit certain symmetries in
the protocol to derive expressions for $P_A^*$ from the above expressions
for $P_B^*$.

Given an $n$-Coin-Game $\G$ define a new $(n+1)$-Coin-Game, $\G'$ by the
rules $\G'_r = 1$, $\G'_{0x}=\G'_{1x}=\G_x$ for $|x|<n$ and
$\G'_{0x}=\G'_{1x}=1-\G_x$ for $|x|=n$. We'd like to argue that the quantum
protocol associated with $\G'$ is equivalent to the protocol associated
with $\G$ but with Alice's and Bob's roles exchanged.

The basic idea is that the first message of $\G'$, which Alice sends to Bob
is the pure state $\ket{0}$. If Bob is cheating this state reveals no extra
information about Alice's state, and if Alice is cheating she has no
incentive to reveal herself as a cheater by sending anything other than the
state $\ket{0}$. The subsequent messages in $\G'$ correspond to those of
$\G$ but with Alice and Bob reversed. The only potential problem
with this argument is that the order of the cheat detection messages (steps
5 and 6) needs to be switched in order to make the protocols
equivalent. However, we argued after formulating the problem as an SDP that
these two steps could be exchanged without increasing or decreasing
$P_B^*$.

Therefore, Bob's maximum probability of winning by cheating in $\G'$, which
we call ${P_B^*}'$ and can be calculated using the above formulas, equals
$P_A^*$. But $\B_r' = \sqrt{\A_r}$ and $\HH_r' = 1-\HH_r$, where the primed
variables are calculated from $\G'$. The conclusion is that
\be
P_A^* = {P_B^*}' = \frac{\B_r'^2}{\HH_r'} = \frac{\A_r}{1-\HH_r}.
\ee

In particular we have proven the main result of this appendix, which is
equivalent to Theorem~\ref{theorem:proto}:
\begin{theorem}
The quantum weak coin-flipping protocol associated to an $n$-Coin-Game $\G$
by Protocol 1 satisfies:
\be
P_A^* =\frac{\A_r}{1-\HH_r}, \qquad P_B^* =\frac{\B_r^2}{\HH_r},
\ee
and $P_A=1-P_B=1-\HH_r$, where $\A$, $\B$ and $\HH$ are defined in terms of
$\G$ by Eqs.~(\ref{eq:h},\ref{eq:ab}).
\end{theorem}

The above result could be made more symmetric between Alice and Bob, if we
were to redefine $\A$ and $\B$ by
\be
\A_x^{(new)} &=& \begin{cases}
\sqrt{\A_x}&\text{$|x|$ even}\cr
\A_x&\text{$|x|$ odd}
\end{cases}\\
\B_x^{(new)} &=& \begin{cases}
\B_x &\text{$|x|$ even}\cr
\sqrt{\B_x} &\text{$|x|$ odd}
\end{cases}
\ee
which could be computed bottom-up by a sequence of linear and
root-mean-squared averages as in Ref.~\cite{me2004}. The new definitions
would also make the conserved quantities such as
Eqs.~(\ref{eq:cons1},\ref{eq:cons2}) have the same expression at even and
odd depths. However, the old definitions make manifest the convexity that
was exploited in the main sections of this paper, and therefore these
definitions were selected.

\section{\label{sec:192}0.192 Revisited}

In this section we shall derive an analytical expression that corresponds
to the bias of $0.192$ found in Ref.~\cite{me2004}. Since the protocol with
bias $0.192$ has been superseded by the results of the present work, we
shall only sketch the proof. Nonetheless, we hope that the techniques used
in deriving this expression, which are rather different to the approach
taken in the rest of the paper, will be of use in some future applications.

The protocols that converged to a bias of $0.192$ had Coin-Games such that
$\G_x=a_{|x|+1}$ for binary nodes. The pattern of zeros and ones on the
leaves was such that, at each depth, the tree $\A_x$ only had two values
which we can call the high value and the low value. The high value only got
updated at even depths whereas the low value only got updated at odd
depths. In particular, the value of the root node could be calculated using
the following sequences: set $H_n=1$ and $L_n=0$ and define
\be
H_k &=& \sqrt{a_{k+1} L_{k+1}^2 + (1-a_{k+1}) H_{k+1}^2},\\
L_k &=& L_{k+1},
\ee
\noindent
for even $k\geq0$, and
\be
H_k &=& H_{k+1},\\
L_k &=& a_{k+1} H_{k+1} +  (1-a_{k+1}) L_{k+1},
\ee
\noindent
for odd $k\geq0$. The value of $A_r$ is then given by $H_0^2$.

The sequence is defined so that $H$ decreases and $L$ increases with
decreasing $k$. At every step the condition $1\geq H_k\geq L_k\geq 0$
holds. For good choices of $a_k$ the two sequences will approach
each other and $H_0$ will be close to $L_0$.

A good sequence of parameters will also have $a_k$ small for large $k$. For
$k$ small, $a_k$ can be larger as long as $a_{k}(H_{k}-L_{k})$
remains small. In such a case, we can use the expansion
\be
H_k &\simeq& H_{k+1} - a_{k+1} \frac{H_{k+1}^2-L_{k+1}^2}{2H_{k+1}},
\ee
\noindent
for even $k$.

Furthermore, if $a_k$ is slowly varying, we can replace it with a continuous
function $a(k)$, and the above computation can be approximated by the
coupled differential equations
\be
\frac{d H}{d k} &=& \frac{a(k)}{2}\, \frac{H^2-L^2}{2H},\\
\frac{d L}{d k} &=& - \frac{a(k)}{2}\, (H-L),
\ee
\noindent
where now $H$ and $L$ are treated as functions of the continuous variable
$k\in[0,n]$. An extra factor of $1/2$ was picked up on the right hand side
of the above equations because $H$ and $L$ only get updated every other
integer in the discrete sequence.

Of course, we are only concerned with the convergence point where $H\simeq
L$. In the limit $n\rightarrow \infty$, and for appropriate $a(k)$, the
two expressions will converge to the same point $H_0=L_0$. To study the
convergence point we can study $H$ as a function of $L$, which satisfies
the differential equation
\be
\frac{d H}{d L}=-\frac{H+L}{2H}.
\ee
\noindent
Surprisingly, the function $a(k)$ drops out of the above expression which
means it only controls the rate of convergence but not the final point of
convergence (assuming it satisfies the requirements discussed above).  In
essence, much the same behavior can be observed by choosing different
$\gamma_n$ sequences for the protocol with bias $1/6$ found in the main
section of this paper.

The differential equation is invariant under simultaneous rescaling of $H$
and $L$, and therefore becomes separable under the change of variables
$H\rightarrow H/L$. Its solutions have the form

\beq
\log\lp( H^2+\frac{1}{2} L H+ \frac{1}{2} L^2\rp) + 
\frac{2}{\sqrt{7}} \arctan\frac{\sqrt{7} L}{4 H +L} = const.
\eeq

The initial condition for the differential equation is $H(L=0)=1$, which
corresponds to the initial starting point when
$k\rightarrow\infty$. Applying the initial condition we obtain $const=0$.
We are interested in the point where $H$ and $L$ converge, that is,
the value $L_0$ such that $H(L_0)=L_0$:
\be
\log 2L_0^2 = - \frac{2}{\sqrt{7}} \arctan\frac{\sqrt{7}}{5}.
\ee
From this value we can obtain $\A_r= L_0^2$. When $a_k$ varies slowly
enough and meets our other requirements we also get $P_A=1/2$ and therefore
$P_A^* = 2L_0^2$. These conditions also guarantee that $P_B^*=P_A^*$, 
hence
\be
P_A^* = P_B^* &=& \Exp\lp[- \frac{2}{\sqrt{7}} \arctan\frac{\sqrt{7}}{5}\rp]
\nonumber\\&\simeq& 0.692181687,
\ee
\noindent
which corresponds to the bias $\epsilon\simeq 0.192$ from Ref.~\cite{me2004}.



\begin{thebibliography}{12}
\expandafter\ifx\csname natexlab\endcsname\relax\def\natexlab#1{#1}\fi
\expandafter\ifx\csname bibnamefont\endcsname\relax
  \def\bibnamefont#1{#1}\fi
\expandafter\ifx\csname bibfnamefont\endcsname\relax
  \def\bibfnamefont#1{#1}\fi
\expandafter\ifx\csname citenamefont\endcsname\relax
  \def\citenamefont#1{#1}\fi
\expandafter\ifx\csname url\endcsname\relax
  \def\url#1{\texttt{#1}}\fi
\expandafter\ifx\csname urlprefix\endcsname\relax\def\urlprefix{URL }\fi
\providecommand{\bibinfo}[2]{#2}
\renewcommand{\eprint}[2][]{\href{#1/#2}{#2}}

\bibitem[{\citenamefont{Mochon}(2004{\natexlab{a}})}]{me2004}
\bibinfo{author}{\bibfnamefont{C.}~\bibnamefont{Mochon}},
  \emph{\bibinfo{title}{Quantum weak coin-flipping with bias of 0.192}}, in
  \emph{\bibinfo{booktitle}{45th Symposium on Foundations of Computer Science
  (FOCS '04)}} (\bibinfo{publisher}{IEEE Computer Society},
  \bibinfo{year}{2004}{\natexlab{a}}), pp. \bibinfo{pages}{2--11},
  \eprint[http://arXiv.org/abs]{quant-ph/0403193}.

\bibitem[{\citenamefont{Spekkens and
  Rudolph}(2002{\natexlab{a}})}]{Spekkens2002}
\bibinfo{author}{\bibfnamefont{R.~W.} \bibnamefont{Spekkens}} \bibnamefont{and}
  \bibinfo{author}{\bibfnamefont{T.}~\bibnamefont{Rudolph}},
  \emph{\bibinfo{title}{Quantum protocol for cheat-sensitive weak coin
  flipping}}, \bibinfo{journal}{Phys. Rev. Lett.}
  \textbf{\bibinfo{volume}{89}}, \bibinfo{pages}{227901}
  (\bibinfo{year}{2002}{\natexlab{a}}),
  \eprint[http://arXiv.org/abs]{quant-ph/0202118}.

\bibitem[{\citenamefont{Ambainis}(2002)}]{Ambainis2002bis}
\bibinfo{author}{\bibfnamefont{A.}~\bibnamefont{Ambainis}},
  \emph{\bibinfo{title}{Lower bound for a class of weak quantum coin flipping
  protocols}} (\bibinfo{year}{2002}),
  \eprint[http://arXiv.org/abs]{quant-ph/0204063}.

\bibitem[{\citenamefont{Ambainis}(2001)}]{Ambainis2002}
\bibinfo{author}{\bibfnamefont{A.}~\bibnamefont{Ambainis}},
  \emph{\bibinfo{title}{A new protocol and lower bounds for quantum coin
  flipping}}, in \emph{\bibinfo{booktitle}{Proceedings on 33rd Annual ACM
  Symposium on Theory of Computing}} (\bibinfo{publisher}{ACM},
  \bibinfo{address}{New York}, \bibinfo{year}{2001}), pp.
  \bibinfo{pages}{134--142}, \eprint[http://arXiv.org/abs]{quant-ph/0204022}.

\bibitem[{\citenamefont{Lo and Chau}(1998)}]{Lo:1998pn}
\bibinfo{author}{\bibfnamefont{H.-K.} \bibnamefont{Lo}} \bibnamefont{and}
  \bibinfo{author}{\bibfnamefont{H.~F.} \bibnamefont{Chau}},
  \emph{\bibinfo{title}{Why quantum bit commitment and ideal quantum coin
  tossing are impossible}}, \bibinfo{journal}{Physica}
  \textbf{\bibinfo{volume}{D120}}, \bibinfo{pages}{177} (\bibinfo{year}{1998}),
\eprint[http://arXiv.org/abs]{quant-ph/9711065}.

\bibitem[{\citenamefont{Goldenberg et~al.}(1999)\citenamefont{Goldenberg,
  Vaidman, and Wiesner}}]{Goldenberg:1998bx}
\bibinfo{author}{\bibfnamefont{L.}~\bibnamefont{Goldenberg}},
  \bibinfo{author}{\bibfnamefont{L.}~\bibnamefont{Vaidman}}, \bibnamefont{and}
  \bibinfo{author}{\bibfnamefont{S.}~\bibnamefont{Wiesner}},
  \emph{\bibinfo{title}{Quantum gambling}}, \bibinfo{journal}{Phys. Rev. Lett.}
  \textbf{\bibinfo{volume}{82}}, \bibinfo{pages}{3356} (\bibinfo{year}{1999}),
\eprint[http://arXiv.org/abs]{quant-ph/9808001}.

\bibitem[{\citenamefont{Spekkens and
  Rudolph}(2002{\natexlab{b}})}]{Spekkens2001}
\bibinfo{author}{\bibfnamefont{R.~W.} \bibnamefont{Spekkens}} \bibnamefont{and}
  \bibinfo{author}{\bibfnamefont{T.}~\bibnamefont{Rudolph}},
  \emph{\bibinfo{title}{Degrees of concealment and bindingness in quantum bit
  commitment protocols}}, \bibinfo{journal}{Phys. Rev. A}
  \textbf{\bibinfo{volume}{65}}, \bibinfo{pages}{012310}
  (\bibinfo{year}{2002}{\natexlab{b}}),
  \eprint[http://arXiv.org/abs]{quant-ph/0106019}.

\bibitem[{\citenamefont{Mochon}(2004{\natexlab{b}})}]{me2003-2}
\bibinfo{author}{\bibfnamefont{C.}~\bibnamefont{Mochon}},
  \emph{\bibinfo{title}{Serial composition of quantum coin-flipping, and bounds
  on cheat detection for bit-commitment}}, \bibinfo{journal}{Phys. Rev.}
  \textbf{\bibinfo{volume}{A70}}, \bibinfo{pages}{032312}
  (\bibinfo{year}{2004}{\natexlab{b}}),
  \eprint[http://arXiv.org/abs]{quant-ph/0311165}.

\bibitem[{\citenamefont{Rudolph and Spekkens}(2004)}]{Spekkens2003}
\bibinfo{author}{\bibfnamefont{T.}~\bibnamefont{Rudolph}} \bibnamefont{and}
  \bibinfo{author}{\bibfnamefont{R.~W.} \bibnamefont{Spekkens}},
  \emph{\bibinfo{title}{Quantum state targeting}}, \bibinfo{journal}{Phys. Rev.
  A} \textbf{\bibinfo{volume}{70}}, \bibinfo{pages}{052306}
  (\bibinfo{year}{2004}), \eprint[http://arXiv.org/abs]{quant-ph/0310060}.

\bibitem[{\citenamefont{Kerenidis and Nayak}(2004)}]{ker-nayak}
\bibinfo{author}{\bibfnamefont{I.}~\bibnamefont{Kerenidis}} \bibnamefont{and}
  \bibinfo{author}{\bibfnamefont{A.}~\bibnamefont{Nayak}},
  \emph{\bibinfo{title}{Weak coin flipping with small bias}},
  \bibinfo{journal}{Inf. Process. Lett.} \textbf{\bibinfo{volume}{89}},
  \bibinfo{pages}{131} (\bibinfo{year}{2004}).

\bibitem[{\citenamefont{Kitaev}()}]{Kitaev}
\bibinfo{author}{\bibfnamefont{A.}~\bibnamefont{Kitaev}},
  \bibinfo{note}{results presented at QIP 2003 (slides and video available from
  MSRI)}.

\bibitem[{\citenamefont{Ambainis et~al.}(2004)\citenamefont{Ambainis, Buhrman,
  Dodis, and Roehrig}}]{Ambainis2003}
\bibinfo{author}{\bibfnamefont{A.}~\bibnamefont{Ambainis}},
  \bibinfo{author}{\bibfnamefont{H.}~\bibnamefont{Buhrman}},
  \bibinfo{author}{\bibfnamefont{Y.}~\bibnamefont{Dodis}}, \bibnamefont{and}
  \bibinfo{author}{\bibfnamefont{H.}~\bibnamefont{Roehrig}},
  \emph{\bibinfo{title}{Multiparty quantum coin flipping}}, in
  \emph{\bibinfo{booktitle}{19th IEEE Annual Conference on Computational
  Complexity}} (\bibinfo{publisher}{IEEE Computer Society},
  \bibinfo{year}{2004}), pp. \bibinfo{pages}{250--259},
  \eprint[http://arXiv.org/abs]{quant-ph/0304112}.

\end{thebibliography}

\end{document}